\documentclass[%
 reprint,
 superscriptaddress,
 amsmath,amssymb,
 aps,
]{revtex4-2}

\usepackage{graphicx}
\usepackage{dcolumn}
\usepackage{bm}
\usepackage{xcolor}

\makeatletter
\providecommand{\hateq}{\mathrel{\mathpalette\my@hat@eq\relax}}
\newcommand{\my@hat@eq}[2]{%
  \begingroup
  \sbox\z@{$\m@th#1=$}%
  \ooalign{%
    \hidewidth\raisebox{-0.3\ht\z@}{$\m@th#1\widehat{}$}\hidewidth\cr
    \box\z@\cr
  }%
  \endgroup
}
\makeatother

\newcommand{\eref}[1]{(\ref{eq:#1})}
\newcommand{\Eref}[1]{Eq.~\eref{#1}}

\newcommand{\SMeso}{\dot{S}_i^{\text{(meso)}}}
\newcommand{\smeso}{\dot{s}_i^{\text{(meso)}}}
\newcommand{\Eqref}[1]{Eq.~\eqref{#1}}

\newcommand{\ave}[1]{\mathchoice{\left\langle #1 \right\rangle}{\langle #1 \rangle}{\langle #1 \rangle}{\langle #1 \rangle}}

\usepackage{xr}
\usepackage{tikz}
\usepackage{mathtools}
\usetikzlibrary{decorations.pathmorphing}
\usetikzlibrary{decorations.markings}
\usetikzlibrary{arrows,shapes,snakes,automata,backgrounds,petri}
\usetikzlibrary{calc}
\usetikzlibrary{patterns}
\usetikzlibrary{decorations.text}
\tikzset{
xxtsubstrate/.style={decorate, 
line width=1pt,
draw=olive, 
decoration=snake, 
segment amplitude=0.75mm, 
line after snake=0.25mm,
line before snake=0.25mm
},
tsubstrate/.style={decorate, 
line width=1pt,
draw=olive, 
decoration=snake, 
segment amplitude=0.5mm, 
segment length=5pt,
segment amplitude=0.2mm, 
line after snake=1mm,
line before snake=1mm
},
Bsubstrate/.style={decorate, 
line width=1pt,
draw=orange, 
decoration=snake,
segment length=5pt,
segment aspect=0,
segment amplitude=0.5mm, 
line after snake=0mm,
line before snake=0mm
},
substrate/.style={decorate, 
line width=1pt,
draw=orange,
decoration=snake, 
segment length=5pt,
segment amplitude=0.5mm, 
line after snake=0.5mm,
line before snake=0.5mm
},
activity/.style={very thick,draw=red,postaction={decorate},
decoration={markings,mark=at position .5 with
{\arrow[draw=red]{>}}}},
tactivity/.style={thick,draw=red,postaction={decorate},
decoration={markings,mark=at position .5 with
{\arrow[draw=red]{>}}}},
tEPSactivity/.style={thick,draw=red,postaction={decorate},
decoration={markings,mark=at position .55 with
{\arrow[draw=red]{>}}}},
tAactivity/.style={thick,draw=red},
Aactivity/.style={very thick,draw=cyan},
Ddensity/.style={very thick,draw=magenta},
tSactivity/.style={thick,draw=red,postaction={decorate},
decoration={markings,mark=at position .7 with
{\arrow[draw=red]{>}}}},
Sactivity/.style={very thick,draw=red,postaction={decorate},
decoration={markings,mark=at position .7 with
{\arrow[draw=red]{>}}}},
polarity/.style={decorate, 
line width=1pt,
draw=red,
decoration={markings,mark=between positions 0 and 1 step 1.1mm with {\draw[red,thick]  (0,0) circle (0.03)} },
segment length=5pt,
segment amplitude=0.5mm, 
},
Bpolarity/.style={decorate, 
line width=1.5pt,
draw=red,
segment length=5pt,
segment amplitude=0.5mm, 
},
density/.style={ 
line width=1.5pt,
draw=magenta,
densely dashed,
segment length=5pt,
segment amplitude=0.5mm, 
},
}

\begin{document}

\preprint{APS/123-QED}

\title{Scaling of Entropy Production under Coarse Graining in Active Disordered Media}

\author{Luca Cocconi}
\affiliation{Department of Mathematics, Imperial College, SW7 2BX London}
\affiliation{Department of Genetics and Evolution, University of Geneva, Geneva}
\affiliation{The Francis Crick Institute, NW1 1AT London}
\author{Guillaume Salbreux}
\affiliation{Department of Genetics and Evolution, University of Geneva, Geneva}
\author{Gunnar Pruessner}
\affiliation{Department of Mathematics, Imperial College, SW7 2BX London}

\date{\today}

\begin{abstract}
Entropy production plays a fundamental role in the study of non-equilibrium systems by offering a quantitative handle on the degree of time-reversal symmetry breaking. It depends crucially on the degree of freedom considered as well as on the scale of description. It was hitherto unknown how the entropy production at one resolution of the degrees of freedom is related to the entropy production at another resolution. This relationship is of particular relevance to coarse grained and continuum descriptions of a given phenomenon. In this work, we derive the scaling of the entropy production under iterative coarse graining on the basis of the correlations of the underlying microscopic transition rates for non-interacting particles in active disordered media. Our approach unveils a natural criterion to distinguish equilibrium-like and genuinely non-equilibrium macroscopic phenomena based on the sign of the scaling exponent of the entropy production per mesostate.
\end{abstract}

\maketitle


\paragraph*{Introduction.}
Under the umbrella term of ``active matter", the study of systems driven by injection and dissipation of energy at the single-agent level has played a prominent role in the development of non-equilibrium physics and expanded its interface with biology. One of the key challenges that arise when developing models of biological matter is to quantify their degree of ``non-equilibriumness", i.e. the extent to which their phenomenology differs from that of a collection of passive particles driven by a bath. The rate of entropy production \cite{seifert_stochastic_2012,cocconi_en} allows for such differentiation by capturing time-reversal symmetry breaking at the microscopic scale \cite{gaspard_time-reversed_2004}. While the injection of energy ensures a strong departure from equilibrium at the single-agent level, these systems do not necessarily exhibit nonequilibrium features at larger spatio-temporal scales \cite{nardini_entropy_2017,shim_macroscopic_2016,egolf_equilibrium_2000,cates_mips,dorosz_entropy_2011}. The question of whether equilibrium is effectively restored at this mesoscopic level requires new methods to quantify how entropy production varies under spatial coarse-graining \cite{esposito_stochastic_2012,amann_communications:_2010,gomez-marin_lower_2008,li2019quantifying,caballero2020stealth,celani2012anomalous,Yu2021inversepower}. We offer a novel perspective on this issue by studying a single-particle driven-diffusion process obtained by perturbing homogeneous diffusion with a non-conservative quenched random forcing. A similar model was recently studied as an effective description for the collective motion of active matter in a random potential \cite{ro_disorder-induced_2020}
and can be seen as a minimal description of a molecular motor self-propelling on a disordered network of cytoskeletal filaments \cite{kahana_active_2008}. 
These models are particular examples of active disordered media
\cite{olsen2021active,peruani2018cold,chepizhko2015active,pincce2016disorder}, 
which we discuss for the first time from a thermodynamic perspective.
On a more abstract level, our model may be seen as a many-particle system randomly exploring a complex phase space. 
From this perspective, our work determines the scaling behaviour of entropy production in a wider class of systems, including biochemical reaction networks \cite{Yu2021inversepower}. 

We first analyse our model on a one-dimensional ring, 
where it is exactly solvable \cite{derrida_velocity_1983}, and we identify a trivial scaling behaviour of the mesoscopic entropy production under block coarse-graining.
We then move to higher dimensional lattices, where we show how the mesoscopic entropy production decays algebraically with block size under block coarse-graining. In order to characterise the non-trivial scaling exponents, we draw on a novel field theoretic formalism based on the Martin-Siggia-Rose construction \cite{zinn2021quantum,hertz_path_2016} and demonstrate that the scaling of the entropy production can be related to the small wavenumber behaviour of the probability current's spectral density by arguments reminiscent of those employed in the treatment of hyperuniform fluctuations \cite{torquato_hyperuniformity_2016}. Our main result, Eq.~\eqref{eq:main_res}, provides a natural criterion to distinguish between equilibrium-like and genuinely non-equilibrium macroscopic phenomena based on the sign of the scaling exponent for the entropy production per mesostate.

\paragraph*{Entropy production and coarse graining.}
The starting point of our analysis is a Markovian jump process satisfying the master equation
\begin{equation}
\label{eq:master_eq}
\dot{P}_n(t) = \sum_{m} (P_m(t) w_{m,n} - P_n(t) w_{n,m})
\end{equation}
for the probability $P_n(t)$, with $w_{n,m}$ the non-negative transition rate from state $n$ to state $m\ne n$. The average rate of internal entropy production at steady-state is defined as \cite{cocconi_en}
\begin{equation}
\dot{S}_i = \frac{1}{2}\sum_{n,m} J_{n,m} \ln \frac{\pi_n w_{n,m}}{ \pi_m w_{m,n}}
\label{eq:si_def}
\end{equation}
where $\pi_n = \lim_{t\to\infty}P_n(t)$ is the steady-state probability mass function and $J_{n,m} = \pi_n w_{n,m} - \pi_m w_{m,n}$ is the net probability current from state $n$ to state $m$. The entropy production $\dot{S}_i$ is non-negative and vanishes for systems that satisfy detailed balance. Computing $\dot{S}_i$ from Eq.~\eqref{eq:si_def} requires complete knowledge of the set of microscopic probability currents $j_{nm} = \pi_n w_{n,m}$, which renders this observable sensitive to time-reversal symmetry breaking induced by energy injection at the microscopic scale. However, this `fully resolved' entropy production might be of little interest in the discussion of effective descriptions at the mesoscopic scale. In recent years, various works have addressed the issue of coarse-graining, which amounts to a partial loss of information about the microscopic currents $j_{nm}$ \cite{esposito_stochastic_2012,amann_communications:_2010,dorosz_entropy_2011,Rahav_2007,Gomez_Marin_2008,nicolis_2011,caballero2020stealth}. The perennial difficulty is that the resulting mesoscopic description is in general non-Markovian \cite{strassberg_negen,bo2017multiple}. 
Esposito \cite{esposito_stochastic_2012} has identified a decomposition of the full entropy production under phase-space partitioning into three non-negative contributions. 
Assuming a separation of timescales between intra-mesostate and inter-mesostate transitions, it was also shown that the mathematical form of the entropy production, \Eqref{eq:si_def}, is recovered at the mesoscopic level.

Previous work has focused on a single coarse-graining step, partly due to constraints of Markovianity. However, if the state space is sufficiently large, it is natural to ask whether such coarse-graining procedure could be performed iteratively, in a spirit similar to Kadanoff's ``block spin'' renormalisation \cite{kadanoff1966}. Characterising the scale dependence of suitable observables such as the entropy production per mesostate
\begin{equation}\label{eq:def_smeso}
\smeso = \SMeso/N^{\rm (meso)}~,
\end{equation}
with $N^{\rm (meso)}$ the number of mesostates at a given coarse-graining level, will then convey important information regarding the degree of activity at different scales. To carry out this programme we first denote the steady-state probability current from mesostate $\alpha$ to mesostate $\beta$ by
\begin{equation}
\label{eq:curr_meso}
j_{\alpha \beta}{(L)} = \sum_{n \in \alpha} \sum_{m \in \beta} \pi_n w_{n,m}
\end{equation} 
with $L$ the characteristic coarse-graining length scale, such that $j_{\alpha \beta}{(1)} = \pi_n w_{nm}$ 
for $\alpha=\{n\}$ and $\beta=\{m\}$.
We then sidestep the issue of Markovianity by postulating an effective mesoscopic entropy production of the form
\begin{equation}
\label{eq:si_meso}
\SMeso(L) = \frac{1}{2}\sum_{\alpha,\beta} (j_{\alpha \beta}{(L)} - j_{\beta \alpha}{(L)}) \ln \frac{j_{\alpha \beta}{(L)}}{j_{\beta \alpha}{(L)}}~,
\end{equation}
as would be measured by any observer faithfully applying \Eref{si_def} to a process observed at a given resolution $L$.
The observable $\SMeso(L)$ should be thought of as the entropy production of a different (Markovian) process constrained to displaying the coarse-grained currents of the original process. A similar approach has been recently discussed to characterise the scaling of energy dissipation in non-equilibrium reaction networks \citep{Yu2021inversepower}.



\begin{figure}
    \includegraphics[scale=1]{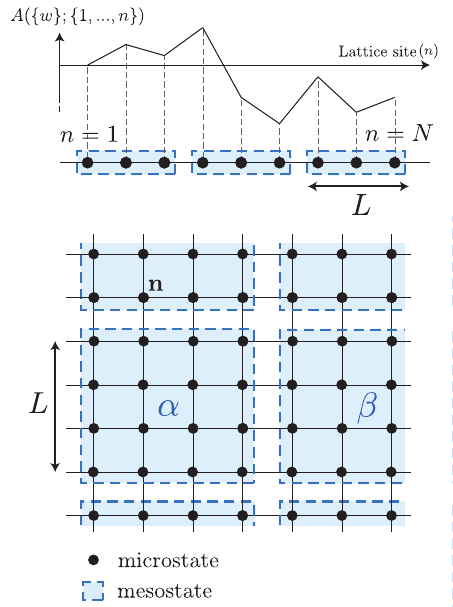}
    \caption{Coarse-graining procedure for a diffusion process on a periodic square lattice perturbed by a spatially quenched, non-conservative disorder. In one dimension (top), the transition rates locally define a random walk with increments $\nu_n = \zeta_{n,n+1}-\zeta_{n+1,n}$ and the affinity $A(\{w\};\{1,..,N\})$ of the closed cycle across all sites is to leading order proportional to the displacement of the random walk after $N$ steps, Suppl.\ Sec.\ \ref{A:derrida}. In higher dimensions (bottom), the random potential picture breaks down locally and the coarse-graining becomes non-trivial due to the current no longer being uniform.} 
    \label{fig:schematic_f1}
\end{figure}

\paragraph*{The model.} 
We will now introduce a minimal driven-diffusion model \cite{SchmittmannZia:1995} on a regular lattice, such that the corresponding non-equilibrium steady-state is amenable to iterative block coarse-graining. Diffusion in a stable potential is the prototypical equilibrium phenomenon but there are many ways to modify the familiar diffusive dynamics into a non-equilibrium process, for example by allowing for an unstable potential \cite{wang_landscape_2015}. An alternative modification, which guarantees a steady-state, is to impose periodic boundary conditions in such a way that a global potential function cannot be defined. To see how this is done, we recall \cite{Schnakenberg1976} that for a Markov jump process, Eq.~\eqref{eq:master_eq}, to have an equilibrium steady state, the affinity
\begin{equation}
\label{eq:affinity_definition}
    A(\{w\};{n_1,\ldots,n_M}) = \ln\frac{w_{n_1, n_2} w_{n_2, n_3} \ ... \ w_{n_M, n_1}}{w_{n_1, n_M} \ ... \ w_{n_3, n_2} w_{n_2, n_1}} ~.
\end{equation}
of every cycle $\{ n_1,n_2,...,n_M,n_1 \}$ has to be exactly zero. Henceforth, we will use the convention that $\ln(0/0)=0$ as some of the $w_{n,m}$ may vanish.
Whenever $A \neq 0$ for any cycle, the system is intrinsically out of equilibrium and we should expect steady-state probability currents \cite{wang_landscape_2015}. A straightforward way of inducing non-equilibrium behaviour is therefore to allow for quenched disorder in the transition rates, which can be interpreted as a non-conservative random forcing \cite{bogachev_random_2006,bouchaud_anomalous_1990}. The resulting disordered steady-state is then non-equilibrium with exceedingly high probability. In the following we therefore consider a homogeneous diffusion process on a periodic lattice in $d$ dimensions and allow for a quenched perturbation to the nearest-neighbour hopping rates such that 
\begin{equation}
\label{eq:perturb_rates}
w_{n,m} = h + \zeta_{n,m} ~,
\end{equation}
with $\zeta_{n,m} > - h$ a set of zero-mean random variables, Fig.~\ref{fig:schematic_f1}. Henceforth, $\langle \cdot \rangle$ will denote averages over this random variable. 

\paragraph*{Ring topology ($d=1$).} The one-dimensional version of this model has previously been considered in the context of random walks in random environments \cite{bouchaud_anomalous_1990,bogachev_random_2006}. Starting from an exact result by Derrida \cite{derrida_velocity_1983} for the net current, $J=J_{n,n+1}$, the entropy production, Eq.~\eqref{eq:si_def}, can be calculated as $\dot{S}_i = J A(\{w\};\{1,2,\ldots,N\})$, 
on the basis of constant $J=\pi_n w_{n,n+1}-\pi_{n+1} w_{n+1,n}$ and the affinity $A$, Eq.~\eqref{eq:affinity_definition}, taken for the cycle passing through all sites of the ring once \cite{Schnakenberg1976}, Suppl.\ Sec.\ \ref{A:derrida}. 
After substituting Eq.~\eqref{eq:perturb_rates} into Eqs.~\eqref{eq:affinity_definition} we eventually obtain
\begin{equation}
\label{eq:st_st_etrprd1d}
    \dot{S}_i = \frac{1}{h N^2} \left( \sum_{n=1}^N \left( \zeta_{n,n+1} - \zeta_{n+1,n} \right) \right)^2 + \mathcal{O}(\zeta^3) ~,
\end{equation}
where $\mathcal{O}(\zeta^k)$ stands for any term proportional to $\zeta_{i_1\pm1,i_1}...\zeta_{i_k\pm1,i_k}$ with any indices $i_1,...,i_k$.
As can be seen by setting $\zeta_{i,j}=0$ for all $i,j$,
the entropy production vanishes at zeroth order in the perturbation, as expected. 
Supp.~Sec.~I explores the weak disorder limit of Eq.~\eqref{eq:st_st_etrprd1d} in more detail.

We now apply the coarse-graining procedure based on Eqs.~\eqref{eq:curr_meso} and \eqref{eq:si_meso}. In one dimension,  the interface between distinct mesostates consists of a single edge (Fig.~\ref{fig:schematic_f1}) and the net current $j_{\alpha\beta}(L)-j_{\beta\alpha}(L) = J$ is independent of the block size $L$. 
The mesoscopic entropy production $\SMeso$, Eq.~\eqref{eq:si_meso}, is thus given by a sum over a subset of the contributions to the microscopic entropy production $\dot{S}_i$, Eq.~\eqref{eq:si_def}. By invoking translational invariance and linearity of expectation, the total entropy production rate in a system with originally $N$ states coarse-grained into mesostate blocks of size $L$ is
\begin{align}
\ave{\SMeso}(L)  &= \left\langle J  \sum_{k=1}^{N/L} \ln \frac{\pi_{kL} w_{kL,kL+1}}{\pi_{kL+1} w_{kL+1,kL}} \right\rangle \\
&= \frac{N}{L} \left\langle J  \ln \frac{\pi_{n} w_{n,n+1}}{\pi_{n+1} w_{n+1,n}}  \right\rangle ~,
\end{align}
with arbitrary state index $n$. 
For uncorrelated noise in the weak disorder limit and using $\pi_n/\pi_m = r_n/r_m$ \cite{derrida_velocity_1983}, $\ave{\SMeso} = 2\lambda/(NhL) + \mathcal{O}(\lambda^2)$, where $\lambda$ denotes the variance of $\zeta_{n,m}$. 
Irrespective of the noise strength, the dependence of the total entropy production on the block size $L$ is solely due to the absence of terms from currents within a microstate block. In one dimension, the entropy production per mesostate $\ave{\smeso}=(L/N)\ave{\SMeso}$, \Eqref{eq:def_smeso}, is thus independent of $L$. No current averaging at interfaces between blocks takes place. 
%
%
%
The situation is qualitatively different in $d > 1$, as we will demonstrate now. 

\begin{figure}
    \includegraphics[scale=0.65]{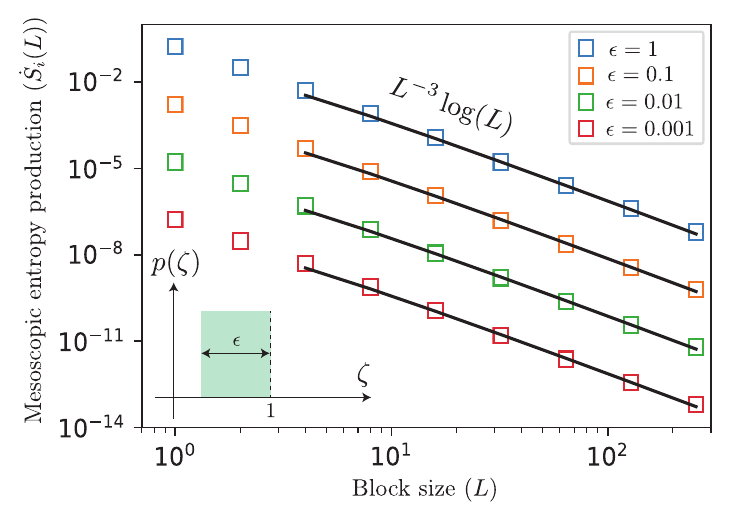}
    \caption{Mesoscopic entropy production as a function of block size and predicted scaling according to Eq.~\eqref{eq:main_res} (case $d=2$,  $\eta=0$) for various noise strengths. The predicted scaling appears to hold numerically beyond the weak disorder approximation. The noise $\zeta_{n,m}$ is taken from a uniform distribution with support $\zeta \in \{[-1,-(1-\epsilon)) \cup (1-\epsilon,1]\}$ and $h=1$ to ensure positivity of the transition rates.
    }
    \label{fig:strong_disorder}
\end{figure}

\paragraph*{Periodic lattices with $d>1$.}
In higher dimensions, the equilibrium condition $A(\{w\};\{n_1,...,n_M\})=0$ is generally broken at the local rather than global scale. No analytical expression for the steady-state currents is available and we resort to a perturbation theory in weak disorder, based on a Martin-Siggia-Rose field theory \cite{zinn2021quantum,hertz_path_2016}, which allows us to extract various static correlation functions in arbitrary dimensions and for a wide class of disorders, 
see Suppl.\ Sec.~\ref{A:ft}. There, we introduce the net microscopic probability current $\bm{J}(\bm{x})=(J^{(1)}(\bm{x}),...,J^{(d)}(\bm{x}))$ as the continuum limit of $J_{\bm{n},\bm{m}}$ on a hypercubic lattice, together with its Fourier transform $\bm{\mathcal{J}}(\bm{k})=(\mathcal{J}^{(1)}(\bm{k}),...,\mathcal{J}^{(d)}(\bm{k}))$. We follow the convention
\begin{equation}
    J^{(a)}(\bm{x}) = \frac{1}{V} \sum_{\bm{k}} \mathcal{J}^{(a)}(\bm{k}) e^{i \bm{k}\cdot \bm{x}} \label{eq:conv_ft}
\end{equation}
with $\bm{k} = 2 \pi \bm{n}/(N\ell)$ ($\bm{n}\in \mathbb{Z}^d$), assuming a hypercubic system, and $V = (N \ell)^d$ the phase space volume, with $\ell$ the dimensionful lattice spacing. 
We assume a disorder characterised by the covariance in Fourier space
\begin{equation}
\label{eq:disord_cov}
\langle \zeta^{(a)}(\bm{k}) \zeta^{(b)}(\bm{k}') \rangle = \tilde{\lambda} |\bm{k}|^{-\eta} \delta_{ab} V \delta_{\bm{k}+\bm{k}',0} 
\end{equation}
for $|\bm{k}| \to 0$, corresponding, for $\eta \neq 0$, to $\langle \zeta^{(a)}(\bm{r}) \zeta^{(b)}(\bm{r}')\rangle \sim \delta_{ab}|\bm{r}-\bm{r'}|^{-d+\eta}$ at $|\bm{r}-\bm{r'}| \to \infty$, where $\zeta^{(a)}$ indicates the disorder affecting edges parallel to the $a$-th dimension of the lattice (see Suppl.\ Sec.~\ref{A:ft} for details of the specification of backward rates). The spectral density tensor of the probability current evaluated at tree level then reads (Suppl.\ Sec.~\ref{A:ft}), for $\bm{k} \neq 0$,
\begin{equation}
\label{eq:sp_dens_j}
\langle \mathcal{J}^{(a)}(\bm{k}) \mathcal{J}^{(b)}(\bm{k}') \rangle = \frac{4 \tilde{\lambda}}{V} \left( \delta_{ab} - \frac{k_a k_b}{|\bm{k}|^2} \right) |\bm{k}|^{-\eta} \delta_{\bm{k}+\bm{k}',0}~.
\end{equation}
Eq.~\eqref{eq:sp_dens_j} matches the general form of the spectral density of a divergence-free, isotropic vector field, which is well known from the theory of turbulence of incompressible fluids \cite{forster_large-distance_1977,monin1975}. For $\eta < 0$, the vanishing of the spectral density as $\bm{k} \to 0$ indicates that the probability current is hyperuniform \cite{torquato_hyperuniformity_2016}, i.e.\  exhibits an anomalous suppression of fluctuations at large wavelengths. The case $\eta=0$ corresponds to uncorrelated (white) noise, $\langle \zeta^{(a)}(\bm{r}) \zeta^{(b)}(\bm{r}')\rangle = \tilde{\lambda} \delta_{ab}\delta(\bm{r}-\bm{r'})$. \\
\begin{figure}
    \includegraphics[scale=0.8]{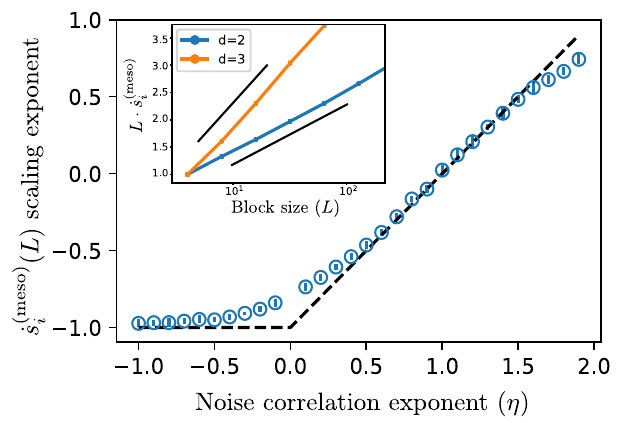}
    \caption{The dependence of the scaling exponent for the entropy production per mesostate, $\dot{s}_i^{\rm(meso)}$, on $\eta$ in the range $-1 \leq \eta \leq d$, shown here for $d=2$ and $N=2048^2$, is well captured by Eq.~\eqref{eq:main_res}, shown in black dashed. For uncorrelated disorder, $\eta=0$, the algebraic scaling of $\dot{s}_i^{\rm (meso)}$ with block size $L$ displays a logarithmic correction, shown in the inset for $d=2,3$, also in agreement with Eq.~\eqref{eq:main_res}. Exact logarithmic scaling is shown in solid black for reference. The ordinate is here normalised to its value for the smallest block size considered. 
    }
    \label{fig:2d_corr}
\end{figure} 
We can now carry out the coarse-graining procedure for the entropy production. First, we note that the mesoscopic entropy production, Eq.~\eqref{eq:si_meso}, is given by a sum over contributions from neighbouring mesostates, $\alpha\ne\beta$ in \eqref{eq:si_meso}. By linearity of expectation, the disorder average $\ave{\SMeso}$ is therefore the expected contribution from a single interface multiplied by the number of interfaces. It follows that, for a hypercubic lattice with coordination number $2d$,
\begin{equation}
\label{eq:ave_simeso}
 \ave{\smeso}  = d \left\langle (j_{\alpha \beta}{(L)} - j_{\beta \alpha}{(L)}) \ln \frac{j_{\alpha \beta}{(L)}}{j_{\beta \alpha}{(L)}} \right\rangle 
\end{equation}
where $\{\alpha,\beta\}$ is any pair of neighbouring mesostates. 
Asymptotically in large $L$, Eq.~\eqref{eq:ave_simeso} can be approximated by
\begin{equation}
\label{eq:scal_si_mes}
 \ave{\smeso} \simeq \frac{d N^d}{h L^{d-1}} \sigma^2_J(L)
\end{equation}
where 
\begin{equation}
\label{eq:def_variance_curr}
\sigma^2_J(L) = \langle(j_{\alpha \beta}{(L)} - j_{\beta \alpha}{(L)})^2 \rangle 
\end{equation}
denotes the variance of the probability current integrated across an interface of linear dimension $L$. This relation is derived in Suppl.\ Sec.~\ref{A:asym_lograt}. In the continuum limit, the asymptotic scaling of the entropy production per mesostate is therefore controlled by the asymptotic variance of the current integrated across the interface between states, which is in turn determined by the small wavenumber behaviour of the current spectral density tensor introduced in Eq.~\eqref{eq:sp_dens_j}. In fact, the relationship between Eqs.~\eqref{eq:sp_dens_j} and \eqref{eq:def_variance_curr} is exactly the type of problem addressed in the study of hyperuniform systems \cite{torquato_hyperuniformity_2016}. Using results from these studies, one obtains
\begin{equation}
\label{eq:var_hyunif}
\sigma^2_J(L) \sim \left\{\begin{array}{lr}
        L^{d-2}, & \text{for } \eta < 0\\
        L^{d-2}\ln(L), & \text{for } \eta = 0\\
        L^{d-2+\eta}, & \text{for } 0<\eta<d
        \end{array}  ~. \right.  
\end{equation}
A more thorough derivation of these scaling laws is provided in Suppl.~ Sec.~\ref{A:hyperu_scaling}. We can think of $\sigma^2_J(L) \sim L^{d-2}$ as scaling with the perimeter of the interface. In this sense, it is instructive to draw a comparison with the case of a non-solenoidal random vector field with spectral density $\langle \mathcal{J}^{(a)}(\bm{k}) \mathcal{J}^{(b)}(\bm{k'})\rangle = (\tilde{\lambda}/V) \delta_{ab} \delta_{\bm{k}+\bm{k}',0}$, in which case the variance of the integrated current instead scales with the area of the interface, $\sigma_J^2(L) \sim L^{d-1}$. The requirement that the steady-state is divergence-free thus plays an important role by imposing long-range correlations in the currents, even when these are not present in the substrate, i.e.\ for $\eta=0$. 
Combining Eqs.~\eqref{eq:scal_si_mes} and \eqref{eq:var_hyunif} we eventually arrive at (Suppl.\ Sec.~\ref{A:hyperu_scaling})
\begin{equation}
\label{eq:main_res}
    \left\langle \dot{s}_i^{\rm (meso)} \right\rangle(L) \propto \frac{\sigma^2_J(L)}{L^{d-1}} \sim \left\{\begin{array}{lr}
        L^{-1}, & \text{for } \eta < 0\\
        L^{-1}\ln(L), & \text{for } \eta = 0\\
        L^{-1+\eta}, & \text{for } 0<\eta<d
        \end{array} \right.~,
\end{equation}
which constitutes our key result. The scaling exponent changes sign at $\eta = 1$, corresponding to $\langle \zeta^{(a)}(\bm{r}) \zeta^{(b)}(\bm{r}')\rangle \sim \delta_{ab}|\bm{r}-\bm{r'}|^{-(d-1)}$, suggesting a quantitative distinction between steady-states that are increasingly ``equilibrium-like'' at larger scales and genuinely non-equilibrium states where dissipation occurs at all scales.
Numerical simulations for $\eta = 0$ in $d=2,3$, as well as the full range $-1 < \eta < d$ in $d=2$ show excellent agreement with our analytical prediction (see Figure \ref{fig:2d_corr}).  Investigating entropy production numerically indicates that this scaling behaviour of the entropy production is unchanged in the strong disorder regime (Figure \ref{fig:strong_disorder}).

\paragraph*{Concluding remarks and outlook.}
Based on the effective entropy production introduced in Eq.~\eqref{eq:si_meso}, we have studied the mesoscopic scaling of the entropy production per mesostate under iterative block coarse graining for a family of non-equilibrium disordered steady-states with generic substrate correlations. We demonstrate 
that coarse-graining of degrees of freedom results in non-trivial \emph{scaling} of the entropy production, Eqs.~\eqref{eq:main_res}, so that its value on one scale is related to its value on another scale in a non-trivial fashion that ultimately draws on the correlations of the transition rates, Eq.~\eqref{eq:disord_cov}.
To characterise our model we have developed a static field-theoretic framework, which complements recent work \cite{antonov_static_2020} by considering problems involving multiplicative noise. 
Our main result, Eq.~\eqref{eq:main_res}, offers  conditions for which active disordered media appear equilibrium-like, $\eta \leq 1$, or genuinely non-equilibrium, $\eta > 1$, at the large scale and in a statistical sense, i.e. when the behaviour is averaged over many realisations. 

\paragraph*{Experiments.} A natural application of the theory above of active disordered media are active particles \cite{bechinger2016active} on irregular surfaces, which phenomenologically behave as randomly driven passive particles \cite{ro_disorder-induced_2020}. This is due to ratchet effects, i.e.\ local asymmetries in the potential driving a net current \cite{reichhardt2017ratchet}. In this case, correlations in the medium can be induced by controlling the ruggedness of the substrate \cite{yu2021engineered}.
Further, in vitro experiments involving active transport by molecular motors in a network of cytoskeletal filaments \cite{salman2005nuclear, bashirzadeh2019encapsulation} often involve cell extracts where these filaments are uniformly disordered ($\eta=0$). Non-trivial, long-range correlations in this type of systems could be induced e.g.\ by coupling tagged filaments to an external magnetic field, as done in \cite{chen2011magnetic} with actin.

\begin{acknowledgments}
LC would like to thank Rosalba Garcia-Millan, Salvatore Torquato, Nikolai Antonov and Polina Kakin for fruitful discussion at various stages of this work. LC and GS acknowledge support from the
Francis Crick Institute, which receives its core funding
from Cancer Research UK (FC001317), the UK Medical
Research Council (FC001317), and the Wellcome Trust
(FC001317).
\end{acknowledgments}

\bibliography{en_scaling_refs}


\end{document}


\preprint{APS/123-QED}

\title{Scaling of Entropy Production under Coarse Graining: Supplementary Material}

\author{Luca Cocconi}
\affiliation{Department of Mathematics, Imperial College, SW7 2BX London}
\affiliation{Department of Genetics and Evolution, University of Geneva, Geneva}
\affiliation{The Francis Crick Institute, NW1 1AT London}
\author{Guillaume Salbreux}
\affiliation{Department of Genetics and Evolution, University of Geneva, Geneva}
\author{Gunnar Pruessner}
\affiliation{Department of Mathematics, Imperial College, SW7 2BX London}

\date{\today}

\maketitle



\beginsupplement
\onecolumngrid

\section{Weak disorder expansion of Derrida's exact result for $d=1$  \label{A:derrida}}
Using an exact result by Derrida \cite{derrida_velocity_1983} the net current $J=J_{n,n+1}$ is homogeneous,
\begin{equation}
\label{eq:derrida_v}
    J = \frac{1}{\sum_{n=1}^N r_n} \left[ 1 - \prod_{m=1}^N \left( \frac{w_{m+1,m}}{w_{m,m+1}}\right)\right]
\end{equation}
for a lattice with $N$ sites, where
\begin{equation}
    r_n = \frac{1}{w_{n,n+1}} \left[ 1 + \sum_{\ell=1}^{N-1} \prod_{j=1}^\ell \frac{w_{n+j,n+j-1}}{w_{n+j,n+j+1}} \right] ~.
\end{equation}
The entropy production, Eq.~\eqref{eq:si_def}, can be calculated as $\dot{S}_i = J A(\{w\};\{1,2,\ldots,N\})$, 
on the basis of constant $J=\pi_n w_{n,n+1}-\pi_{n+1} w_{n+1,n}$ and the affinity $A$, Eq.~\eqref{eq:affinity_definition}, taken for the cycle passing through all sites of the ring once \cite{Schnakenberg1976}. 
Eq.~\eqref{eq:perturb_rates} can be substituted into Eqs.~\eqref{eq:affinity_definition} and \eqref{eq:derrida_v} to obtain
\begin{equation}
\label{eq:st_st_flux1d}
    J = \frac{hA}{N^2} + \mathcal{O}(\zeta^2) = \frac{1}{N^2}\sum_{n=1}^N \left( \zeta_{n,n+1} - \zeta_{n+1,n}\right) + \mathcal{O}(\zeta^2) ~.
\end{equation}
Correspondingly, using $\dot{S}_i = J A$,
\begin{equation} \label{eq:en_suppl}
    \dot{S}_i = \frac{1}{h N^2} \left( \sum_{n=1}^N \left( \zeta_{n,n+1} - \zeta_{n+1,n} \right) \right)^2 + \mathcal{O}(\zeta^3) ~,
\end{equation}
where $\mathcal{O}(\zeta^k)$ stands for any term proportional to $\zeta_{i_1\pm1,i_1}...\zeta_{i_k\pm1,i_k}$ with any indices $i_1,...,i_k$. This is the expression we refer to in the main text, Eq.~\eqref{eq:st_st_etrprd1d}.
In this regime, the current is dominated by the average of the approximate potential gradients, $(\zeta_{n,n+1}-\zeta_{n+1,n})/h$.
The physical picture provided by Eq.~\eqref{eq:st_st_flux1d} shows that the weak disorder expansion, $|\zeta_{n,m}| \ll h$, is equivalent to a linear response theory, $J \propto A$, with the cycle affinity playing the role of the thermodynamic force \cite{groot_non-equilibrium_2003}. 
The behaviour away from the weak disorder limit is more subtle. In particular, the current $J$ and therefore $\dot{S}_i$ are no longer a function only of the affinity but depend on the whole random potential profile. This is due to the presence of traps, i.e. local minima of the random potential, at which probability tends to concentrate \cite{compte_localisation_1998}.

\begin{figure}
    \includegraphics[scale=0.85]{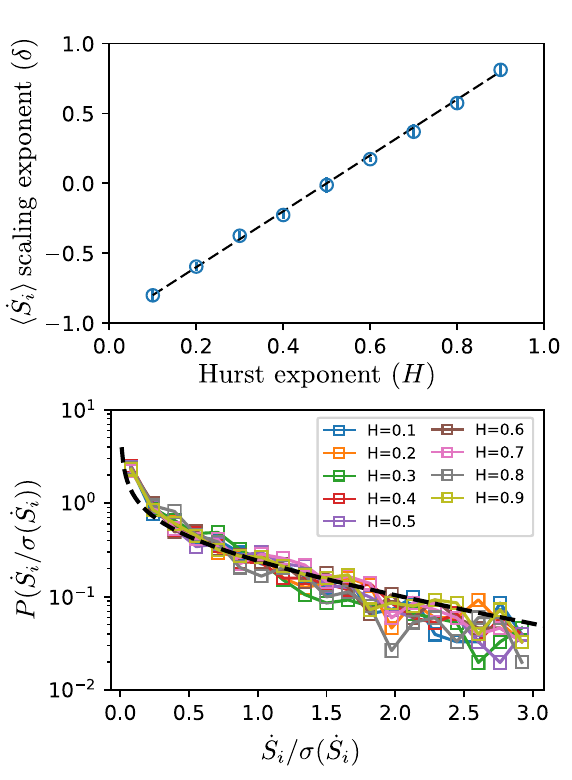}
    \caption{In one dimension and for weak disorder, the scaling of the entropy production with system size is controlled by the Hurst exponent according to Eq.~\eqref{eq:hurst_scaling}. Top: the exponent $\entropyScalingExponent$ obtained by fitting numerical estimates of $\langle \dot{S}_i \rangle$ against $N^\entropyScalingExponent$ for different Hurst exponents $H \in (0,1)$, shown with error bar in blue, is in perfect agreement with the analytical prediction, 
    $\entropyScalingExponent=2H-2$, Eq.~\eqref{eq:hurst_scaling}, shown as a black dashed line. Bottom: the distribution of the entropy production $\dot{S}_i$ as a random variable is well described by the chi-squared distribution, shown as a black dashed line, as predicted by the theory.}
    \label{fig:1d_scaling}
\end{figure}

So far, we have not made any assumption about the autocorrelation of the noise $\zeta$. However, by relating \Eqref{eq:en_suppl} to the square of the distance travelled by a random walker with zero-mean increments $\nu_n= \zeta_{n,n+1} - \zeta_{n+1,n}$, Fig.~\ref{fig:schematic_f1}, we conclude that the system-size scaling of the disorder expectation $\langle \dot{S}_i \rangle $ in the large $N$ limit is controlled by the Hurst exponent $H \in (0,1)$ \cite{mandelbrot_fbm} of the random walk. In particular,
\begin{equation}
\label{eq:hurst_scaling}
\langle \dot{S}_i \rangle \sim N^\entropyScalingExponent = N^{2H-2}~,
\end{equation}
where $H=1/2$ corresponds to uncorrelated (or short-range correlated) disorder. 
By the central limit theorem for sufficiently uncorrelated $\nu_n$, and more generally when $\sum_{n=1}^N \nu_n$ is a Gaussian random variable, the affinity is also Gaussian.
As $\dot{S}_i$, \Eqref{eq:en_suppl}, is essentially the square of the affinity, its distribution is of the chi-squared type. These results are indeed confirmed by simulations, Fig.~\ref{fig:1d_scaling}. 

\section{A Static Path Integral Approach to Diffusion with Quenched Noise \label{A:ft}}

In the following we develop the field theoretic formalism used to study the correlation function of the steady-state probability currents for the model of diffusion on a lattice with spatially quenched, non-conservative noise in the hopping rates. This formalism is based on the well-known response-field construction by Martin-Siggia-Rose (MSR) and Janssen-De Dominicis \cite{zinn2021quantum,hertz_path_2016} but extends it in two ways: first, our approach deals with multiplicative noise in a static framework, complementing recent work by Antonov and co-workers \cite{antonov_static_2020}, which focused on the case of additive noise. 
Second, we show how the MSR `trick' of imposing physically relevant relations between the noise field and physical observables by means of suitable resolutions of identity can be exploited to directly probe observables that depend explicitly on the noise (in this case, the probability current). The formalism is first developed in one dimension, $d=1$, for the case of white noise and it is subsequently generalised to higher integer dimensions and correlated noise. It turns out that we ultimately do not need the full field theory nor the renormalisation group, because the upper critical dimension $d_c=2$ coincides with the lowest dimension we draw on the field theory. Nevertheless, the framework provides us with a firm basis to reliably determine the upper critical dimension and will form the well-tested foundation for the analysis of more advanced problems.\\

The analytical expressions derived in the following are in agreement with numerical results, as shown in Fig.~\ref{fig:1d_scaling} for the one-dimensional problem and in Figs.~\ref{fig:2d_corr} and \ref{fig:compare_symm} for its generalisation to higher dimensional lattices. In one dimension, the steady state probability mass function $\pi_n$ for a given noise realisation $\{\zeta_{ m,m \pm 1 }\}$ is solved exactly by means of Kirchhoff's theorem \cite{Schnakenberg1976} and the homogeneous current is obtained straightforwardly as $J = \pi_n w_{n,n+1} - \pi_{n+1} w_{n_1,n}$ for an arbitrary $n$. The entropy production is then calculated as $\dot{S}_i = J A$ on the basis of Eq.~\eqref{eq:affinity_definition}. In higher dimensions, a numerical approximation to the steady state probability is instead obtained by evolving a homogeneous initial condition, $P_n(t=0) = N^{-d}$, according to the master equation, Eq.~\eqref{eq:master_eq}. 

\subsection{One dimensional case, uncorrelated noise.}
\begin{figure}
    \centering
    \includegraphics[scale=0.8]{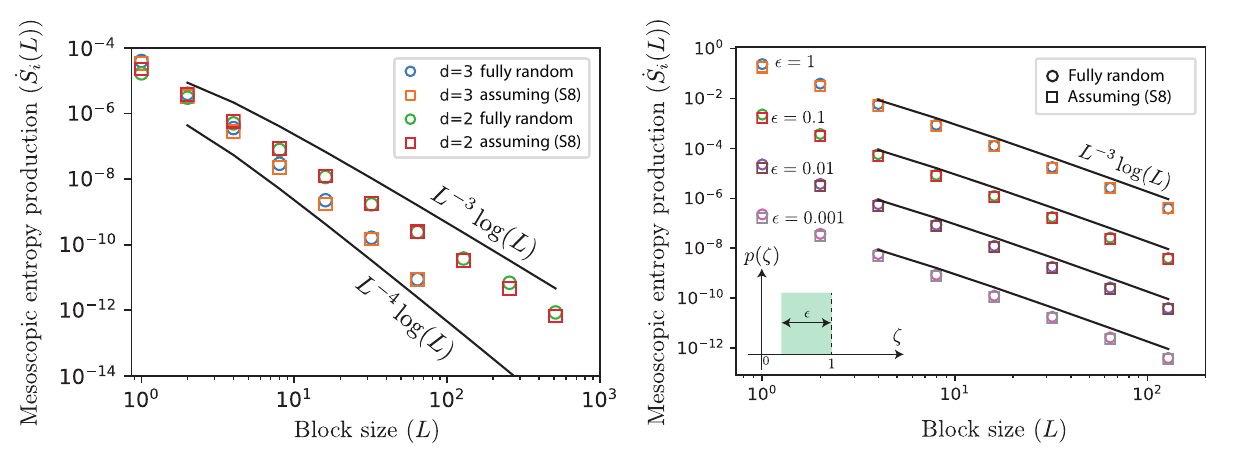}
    \caption{The scaling behaviour of the mesoscopic entropy production $\dot{S}_i(L)$, Eq.~\eqref{eq:si_meso}, as a function of block size $L$ is not modified by imposing the antisymmetric condition \eqref{eq:antsymm_assmp} on the transition rates, as shown for $d=2,3$ in the case of uncorrelated disorder, $\eta=0$ (left panel). 
Moreover, while out analytical approach relies on a weak disorder limit to allow for rate perturbation to be Gaussian without running into unphysical negative rates, numerical experiments suggest that the scaling laws we obtain remain valid in the strong disorder regime (independent of whether \eqref{eq:antsymm_assmp} is imposed), which is explored by considering a noise on the transition rates that is homogeneously distributed in the range $|\zeta| \in (1-\epsilon,1]$ for various choices of the variance $\epsilon$ (right panel for $d=2$ and $\eta = 0$). Theoretical predictions for the asymptotic scaling behaviour are plotted for reference (black curves in both panels).}
    \label{fig:compare_symm}
\end{figure}

We consider a one-dimensional diffusion process on a ring of size $N$ characterised by a homogeneous hopping rate $h$ and a quenched perturbation $\zeta_{i , i \pm 1}$ to the hopping rate from site $i$ to site $i \pm 1$ satisfying $\zeta_{i , i \pm 1} + h >0$. We denote by $\phi_i$ the steady-state probability mass function at site $i$, with $i=1,2,\cdots,N$. By normalisation, $\sum_i \phi_i = 1$. In the following we apply periodic boundary conditions to all indices $i$, so that $i=0$ is equivalent to $i=N$ and $i=N+1$ is equivalent to $i=1$. The steady-state probability mass function is determined implicitly by the master equation
\begin{align}
    0 = \partial_t \phi_i &= h(\phi_{i-1} + \phi_{i+1}-2\phi_i) + \zeta_{i-1 , i} \phi_{i-1} - \zeta_{i , i-1} \phi_i - \zeta_{i , i+1} \phi_i + \zeta_{i+1 , i} \phi_{i+1} \nonumber \\
    &= h (\phi_{i-1} + \phi_{i+1}-2\phi_i) + (\delta_{i-1}\phi_{i-1} + \delta_{i+1}\phi_{i+1}-2\delta_{i}\phi_i) + (\zeta_{i-1} \phi_{i-1} - \zeta_{i+1} \phi_{i+1})  \label{ss_def_int} 
\end{align}
where for the second equality we have redefined the perturbation to the hopping rates according to
\begin{equation}
    \zeta_{i,i+1} = \delta_i + \zeta_i \quad \text{and} \quad \zeta_{i,i-1} = \delta_{i} - \zeta_i ~.
\end{equation}
The right hand side of Eq.~\eqref{ss_def_int} can be written more suggestively in continuum notation as a drift-diffusion equation,
\begin{equation}
    0 = \partial_t \phi = \partial_x[(h + \delta) \partial_x \phi] + \partial_x[( \partial_x\delta -\zeta ) \phi]~,
\end{equation}
where the symmetric part of the quenched noise, denoted by $\delta_i$, appears both as an inhomogeneous perturbation of the diffusion constant $h$ and, in the drift term, as a random forcing. The antisymmetric part of the noise, $\zeta_i$, on the other hand, only affects the drift term.
For simplicity, we will henceforth assume that the perturbation to the transition rate matrix satisfies
\begin{equation}
\label{eq:antsymm_assmp}
     \zeta_{i , i+1} = - \zeta_{i , i-1} = \zeta_{i} \quad \text{i.e.} \quad \delta_i = 0 ~.
\end{equation}
The lattice master equation \eqref{ss_def_int} thus reduces to
\begin{equation}
    0 = \partial_t \phi_i = h(\phi_{i-1} + \phi_{i+1}-2\phi_i) + \zeta_{i-1} \phi_{i-1} - \zeta_{i+1} \phi_{i+1} \label{ss_def}
\end{equation}
Numerical investigation indicates that this assumption does not modify the scaling behaviour of the mesoscopic entropy production, Figure \ref{fig:compare_symm}. For weak disorder, this observation can be rationalised by arguing that the diffusive behaviour is dominated by its homogeneous component, while the drift term is dominated at large spatial scales by the term of lowest order in the spatial derivatives. Importantly, for i.i.d.\ rates, the steady-state remains generically non-equilibrium, which can be verified by computing the cycle affinity, Eq. \eqref{eq:affinity_definition}, and exhibits non-zero steady-state currents. 
We further define the probability current field 
\begin{align}
\label{current_def}
    J_{i,i+1}  = (h + \zeta_{i,i+1})\phi_i - (h+\zeta_{i+1,i})\phi_{i+1} = h(\phi_i - \phi_{i+1}) + \zeta_i \phi_i + \zeta_{i+1} \phi_{i+1} ~.
\end{align}
as the \emph{net} current from site $i$ to site $i+1$, such that
\begin{equation}
    \partial_t \phi_i = -(J_{i,i+1} - J_{i-1,i})~,
\end{equation}
and introduce a symmetrised, local, average current 
\begin{equation}
 J_i = \frac{1}{2}\left( J_{i,i+1} + J_{i-1,i}\right) = \frac{1}{2} [ h(\phi_{i-1} - \phi_{i+1}) + \zeta_{i-1} \phi_{i-1 } + \zeta_{i+1} \phi_{i+1 } + 2 \zeta_{i} \phi_{i} ] ~.
 \label{eq:curr_mean_def}
\end{equation}
In one dimension at stationarity, $J_{i,i+1}=J_{i-1,i}=J_i$ and there is therefore no difference between the local averaged currents and the more microscopic ones. 

The non-negativity of the transition rates means that $\zeta_i + h \geq 0$. However, in order to be able to cast Eqs.~\eqref{ss_def} and \eqref{current_def} into a path integral form, we will assume that $\zeta_{i}$ is a weak Gaussian noise with variance $\lambda$ and probability
\begin{equation}
\label{gaussian_prob}
    P[\zeta_{i}] \propto  {\rm exp}\left( \sum_{i=1}^N  -\frac{\zeta_{i}^2}{2 \lambda} \right) ~.
\end{equation}
This assumption is justified \emph{a posteriori} by comparing the analytical predictions in the weak disorder limit with numerical simulations where the positivity of the transition rates is confirmed at initialisation.
Introducing the short-hand notation for the functional integral measure
\begin{equation}
    \mathcal{D}\zeta = \prod_i d \zeta_{i}~,
\end{equation}
the expectation of a general noise-dependent observable $\mathcal{O}[\zeta]$ can be written as the path integral
\begin{equation}
\label{eq:expect_int}
    \langle \mathcal{O} \rangle = \int \mathcal{D}\zeta \ \mathcal{O}[\zeta] P[\zeta]~.
\end{equation}
To impose the steady-state Eq.~\eqref{ss_def} at every site, we define 
\begin{equation}
    f_i(\phi_{i-1},\phi_i,\phi_{i+1}) = h(\phi_{i-1} + \phi_{i+1}-2\phi_i) + \zeta_{i-1} \phi_{i-1} - \zeta_{i+1} \phi_{i+1}
\end{equation}
and introduce the following resolution of identity written in terms of Dirac delta functions fixing the functional relation between the probability mass function $\phi_i$ and the noise $\zeta_i$
\begin{align}
    1 &= \int \mathcal{D}f \prod_{i=1}^n \delta(f_i) \\
    &= \mathcal{M}[\zeta] \int \mathcal{D}\phi \ \prod_{i=1}^N \delta \Big( h(\phi_{i-1} + \phi_{i+1}-2\phi_i) + \zeta_{i-1} \phi_{i-1} - \zeta_{i+1} \phi_{i+1} \Big) \label{eq:res_st2} \\
    &= \mathcal{M}[\zeta] \int \mathcal{D}\phi \int \mathcal{D}\tilde{\phi} \ {\rm exp}\left[ - \sum_{i=1}^N \tilde{\phi}_i \Big( - h(\phi_{i-1} + \phi_{i+1}-2\phi_i) - \zeta_{i-1} \phi_{i-1} + \zeta_{i+1} \phi_{i+1} \Big) \right] \label{identity_1}
\end{align}
with $\mathcal{D}f = df_1...df_N$, $\mathcal{D}\tilde{f} = d\tilde{f}_1...d\tilde{f}_N/(2\pi)^N$ and $\mathcal{M}[\zeta] = |d f_i/d \phi_j|$ the Jacobian of the transformation $f \to \phi$. 
To go from Eq.~\eqref{eq:res_st2} to Eq.~\eqref{identity_1} we have used the Fourier representation of the Dirac delta function,
\begin{equation}
    \delta (f_i) = \int \frac{d\tilde{\phi}_i}{2\pi} \ {\rm exp}\left[ - \tilde{\phi}_i f_i \right]
\end{equation}
where $\tilde{\phi}_i$ is a purely imaginary auxiliary field, sometimes referred to as the response field.
While $\mathcal{M}$ is independent of $\phi$ for a general master equation and can thus be taken outside of the $\phi$ integral, the Jacobian retains a dependence on the random variables $\zeta_1,...,\zeta_N$ due to the multiplicative nature of the latter. While the sign of the auxiliary field determining that of the exponent in Eq.~\eqref{identity_1} is arbitrary, our choice gives the conventional sign for the response propagator \cite{hertz_path_2016}. 
The determinant $\mathcal{M}[\zeta]$ can be computed by means of Faddeev-Popov ghosts \cite{zinn2021quantum,hertz_path_2016}, a standard procedure which requires the introducing two new Grassmann fields, denoted $\xi_i$ and $\bar{\xi}_i$, satisfying the anti-commutation relations $\{\xi_i,\xi_j \} = \{\bar{\xi}_i,\bar{\xi}_j \} =\{\bar{\xi}_i,\xi_j \} = 0$. Using the Grassmann path integral representation of the determinant \cite{hertz_path_2016},
\begin{align}
    \mathcal{M}[\zeta] &= \int \mathcal{D}[\xi,\bar{\xi}] \ {\rm exp} \left[ \sum_{ij} \bar{\xi}_i \left( \frac{df_i}{d\phi_j} \right) \xi_j \right] \\
    &= \int \mathcal{D}[\xi,\bar{\xi}] \ {\rm exp} \left[ \sum_{i} \bar{\xi}_i \Big( h(\xi_{i-1} + \xi_{i+1}-2\xi_i) + \zeta_{i-1} \xi_{i-1} - \zeta_{i+1} \xi_{i+1} \Big)  \right] \label{eq:ghost_jacobian}~.
\end{align}
This path integral has the same structure as that in Eq.~\eqref{identity_1} because differentiating $f_i$ with respect to $\phi_j$ effectively returns the Markov matrix of the process, now acting to the right on the Grassmann field $\xi_j$. The Jacobian \eqref{eq:ghost_jacobian} couples the Faddeev-Popov ghosts $\xi_i$ and $\bar{\xi}_i$ to the noise field $\zeta_i$, which is in turn coupled to the probability mass function $\phi_i$ and its response $\tilde{\phi_i}$ via Eq.~\eqref{identity_1}. Integrating out the noise will thus result in a coupling between the ghosts and the MSR fields, as we will demonstrate after introducing the current field $J_i$ and its response $\tilde{J}_i$. \\
In principle, the current $J_i$ is fully defined in Eq.~\eqref{eq:curr_mean_def}. However, this definition ceases to be useful once the noise field $\zeta_i$ has been integrated out, rendering $J_i$ intractable. To avoid this, we introduce a resolution of the identity similar to Eq.~\eqref{identity_1} to enforce the definition Eq.~\eqref{eq:curr_mean_def} of the probability current $J_{i}$. It reads
\begin{align}
    1 &= \int \mathcal{D}J \ \prod_{i=1}^N \delta \Big( J_{i} - \frac{1}{2} [ h(\phi_{i-1} - \phi_{i+1}) + \zeta_{i-1} \phi_{i-1 } + \zeta_{i+1} \phi_{i+1 } + 2 \zeta_{i} \phi_{i} ] \Big) \\
    &= \int \mathcal{D}J \int \mathcal{D}\tilde{J} \ {\rm exp}\left[ - \sum_{i=1}^N \tilde{J}_{i} \Big( J_{i} - \frac{1}{2} [ h(\phi_{i-1} - \phi_{i+1}) + \zeta_{i-1} \phi_{i-1 } + \zeta_{i+1} \phi_{i+1 } + 2 \zeta_{i} \phi_{i} ] \Big) \right] \label{identity_2}~.
\end{align}
where the trivial Jacobian $\mathcal{M}\equiv 1$ has been omitted. 

Inserting Eqs.~\eqref{identity_1}, \eqref{eq:ghost_jacobian} and \eqref{identity_2} into the Gaussian path integral \eqref{eq:expect_int} we obtain
\begin{align}
    \langle \mathcal{O}[\phi,J] \rangle &\propto \int \mathcal{D}\phi \int \mathcal{D}\tilde{\phi} \int \mathcal{D}J \int \mathcal{D}\tilde{J} 
    \int \mathcal{D}\xi \int \mathcal{D}\bar{\xi}
    \int \mathcal{D}\zeta \nonumber \\
    &\ {\rm exp}\left[ - \sum_{i=1}^N \tilde{\phi}_i \Big( - h(\phi_{i-1} + \phi_{i+1}-2\phi_i) - \zeta_{i-1} \phi_{i-1} + \zeta_{i+1} \phi_{i+1} \Big) \right] \nonumber \\
    \times &\ {\rm exp}\left[ - \sum_{i=1}^N \tilde{J}_{i} \Big( J_{i} + \frac{h}{2}(\phi_{i+1} - \phi_{i-1}) - \frac{1}{2}(\zeta_{i-1}\phi_{i-1} + \zeta_{i+1} \phi_{i+1} +2 \zeta_i \phi_i) \Big) \right] \nonumber \\
    \times &\ {\rm exp} \left[ \sum_{i} \bar{\xi}_i \Big( h(\xi_{i-1} + \xi_{i+1}-2\xi_i) + \zeta_{i-1} \xi_{i-1} - \zeta_{i+1} \xi_{i+1} \Big)  \right] {\rm exp}\left( \sum_{i=1}^N  -\frac{\zeta_{i}^2}{2\lambda} \right) \mathcal{O}[\phi,J] \label{full_expectation_MSR} ~.
\end{align}
Since the noise $\zeta$ enters the exponent only linearly or quadratically, the functional integral over $\zeta$ is Gaussian and can be performed in closed form. In particular, isolating all terms on the right-hand side of Eq.~\eqref{full_expectation_MSR} that depend on $\zeta$,
\begin{align}
    \int \mathcal{D}\zeta \ &{\rm exp}\left( \sum_{i=1}^N  -\frac{\zeta_{i}^2}{2\lambda} + \tilde{\phi}_i (\zeta_{i-1}\phi_{i-1} - \zeta_{i+1} \phi_{i+1}) + \frac{\tilde{J}_{i}}{2} (2 \zeta_i \phi_i + \zeta_{i+1}\phi_{i+1}+ \zeta_{i-1}\phi_{i-1}) + \bar{\xi}_i (\zeta_{i-1}\xi_{i-1} - \zeta_{i+1}\xi_{i+1}) \right) \label{noise_integr} \\
    = \int \mathcal{D}\zeta \ &{\rm exp}\left( \sum_{i=1}^N  -\frac{\zeta_{i}^2}{2\lambda} + \zeta_i \Big[(\tilde{\phi}_{i+1}-\tilde{\phi}_{i-1})\phi_i + \frac{1}{2} (2\tilde{J}_{i} + \tilde{J}_{i+1} + \tilde{J}_{i-1})\phi_i + (\bar{\xi}_{i+1}-\bar{\xi}_{i-1})\xi_i \Big] \right) \\
    \propto \ &{\rm exp}\left( \frac{\lambda}{2} \sum_{i=1}^N  \Big[(\tilde{\phi}_{i+1}-\tilde{\phi}_{i-1})\phi_i + \frac{1}{2}(2\tilde{J}_{i} + \tilde{J}_{i+1} + \tilde{J}_{i-1})\phi_i + (\bar{\xi}_{i+1}-\bar{\xi}_{i-1})\xi_i \Big]^2 \right) ~.
\end{align} 
In the first equality we have re-indexed the summand, applying periodic boundary conditions, so that the noise field always appears with the same index and the completion of squares producing the last line can be carried out straightforwardly. We can now rewrite Eq.~\eqref{full_expectation_MSR} as
\begin{align}
    \langle \mathcal{O}[\phi,J] \rangle &= \int \mathcal{D} [\phi, \tilde{\phi}, J, \tilde{J},\xi,\bar{\xi}] \ {\rm exp}\left[ - (\mathcal{A}_0 + \mathcal{A}_{\rm int}) \right] \mathcal{O}[\phi,J]
\end{align}
where we have split the action into its bilinear part, denoted $\mathcal{A}_0$, and an interacting part, denoted $\mathcal{A}_{\rm int}$, which includes all higher order terms in products of fields. Denoting by
\begin{equation}
\label{eq:discr_1d}
    \partial_d \phi_i = \frac{1}{2}(\phi_{i+1} - \phi_{i-1})
\end{equation}
and 
\begin{equation}
\label{eq:discr_2d}
    \partial_d^2 \phi_i = \phi_{i+1} - 2\phi_{i} + \phi_{i-1}
\end{equation}
the discrete first and second spatial derivatives, respectively, these contributions are given explicitly by
\begin{equation}
    \mathcal{A}_0 = \sum_{i=1}^N -h \tilde{\phi}_i \partial_d^2 \phi_i -h \bar{\xi}_i \partial_d^2 \xi_i + \tilde{J}_{i} J_{i} + h \tilde{J}_{i} \partial_d \phi_i 
    \label{eq:bilinear_action}
\end{equation}
and
\begin{alignat}{2}
    \mathcal{A}_{\rm int} 
    &=  - 2\lambda &&\sum_{i=1}^N \left[(\partial_d \tilde{\phi}_i)^2 \phi_i^2 + \tilde{J}_i^2 \phi_i^2 + 2 \tilde{J}_i (\partial_d \tilde{\phi}_i) \phi_i^2 + 2 \tilde{J}_i (\partial_d \bar{\xi}_i) \phi_i \xi_i + 2 (\partial_d \tilde{\phi}_i) (\partial_d\bar{\xi}_i)\phi_i \xi_i \right] + \rm{h.o.t.} 
    \label{eq:int_action}
\end{alignat}
where h.o.t.\
denotes higher order terms in the spatial derivatives. In the following we are interested in the large scale (small wavenumber) features of the density and current correlation function, and we will thus ignore contributions from h.o.t.\ on relevance grounds.
To arrive at Eq.~\eqref{eq:int_action} we have used the anticommutation relation $\{ \xi_i,\xi_j\} = 0 $, implying $\xi_i^2 = 0$, to eliminate the quartic term in the ghost fields.  \\

Perturbative calculations by means of Feynman diagrams are more easily carried out in Fourier representation. For this reason we introduce the discrete Fourier transform of the fields $\{ \phi, \tilde{\phi}, J, \tilde{J}, \xi, \bar{\xi} \}$, which we denote $\{ \Phi, \tilde{\Phi}, \mathcal{J}, \tilde{\mathcal{J}}, \Xi, \bar{\Xi} \}$, respectively, according to the convention
\begin{equation}
    \phi_j = \frac{1}{N} \sum_{n=0}^{N-1} e^{2\pi i j n/N} \Phi_n \quad \text{and}\quad N \delta_{n,m} = \sum_{j=0}^{N-1} e^{2\pi i j (n-m)/N }~,
\end{equation}
whence
\begin{equation}
    \Phi_n = \sum_{j=0}^{N-1} e^{-2\pi i j n/N} \phi_j ~.
\end{equation}
We proceed by introducing Fourier-summed fields in the non-linear action. A typical interaction term in the Fourier representation of the action reads
\begin{equation}
     \lambda \sum_{i=1}^N (\partial_d \tilde{\phi}_i)^2 \phi_i^2  = - \frac{ \lambda}{N^4} \sum_{k_1,k_2,k_3,k_4} 
     \sin(k_1)\sin(k_2) 
     \tilde{\Phi}_{k_1} \tilde{\Phi}_{k_2} \Phi_{k_3} \Phi_{k_4} N \delta_{k_1+k_2+k_3+k_4,0}~. \label{eq:exmpl_ft_int} ~. \\
\end{equation}

The field $\phi_i$ represents a probability mass function and it is therefore natural to study fluctuations around the normalised homogeneous steady-state $\langle \phi_i \rangle = N^{-1}$, so that $\Phi_{k=0}=1$. 
Instead of implementing $\Phi_{k=0}=1$ through the action along the lines of Eqs.~\eqref{ss_def} and \eqref{eq:curr_mean_def}, we perform a 
change of variable $\phi_i \to N^{-1} + \phi_i$ at the level of Eqs. \eqref{eq:bilinear_action} and \eqref{eq:int_action}, which leaves the bilinear action $\mathcal{A}_0$ unchanged but generates a number of new terms in $\mathcal{A}_{\rm int}$. For example,
\begin{equation}
 \gamma (\partial_d \tilde{\phi}_i)^2 \phi_i^2 \to \gamma (\partial_d \tilde{\phi}_i)^2 \phi_i^2 +  \gamma' (\partial_d \tilde{\phi}_i)^2 \phi_i + \gamma'' (\partial_d \tilde{\phi}_i)^2 ~,
\end{equation}
with $\gamma = \lambda$, $\gamma' = 2 N^{-1}\lambda$ and $\gamma'' = N^{-2} \lambda$ at bare level. 
The full shifted action reads, again at bare level,
\begin{align}
    \mathcal{A}_{\rm int}
    =  - 2\lambda\sum_{i=1}^N \Big[ &(\partial_d \tilde{\phi}_i)^2 (\phi_i + N^{-1})^2 + \tilde{J}_i^2 (\phi_i + N^{-1})^2 + 2 \tilde{J}_i (\partial_d \tilde{\phi}_i) (\phi_i + N^{-1})_i^2 \nonumber \\
    &+  2 \tilde{J}_i (\partial_d \bar{\xi}_i) (\phi_i + N^{-1}) \xi_i + 2 (\partial_d \tilde{\phi}_i) (\partial_d\bar{\xi}_i)(\phi_i + N^{-1}) \xi_i \Big] + \rm{h.o.t.} 
    \label{eq:int_action_shift}
\end{align}
The various coupling are more easily identifiable in the diagrammatic notation that we shall introduce shortly.

\subsection{Propagators and diagrammatics}
The bare propagators of the theory can then be extracted directly from the bilinear action, Eq.~\eqref{eq:bilinear_action}. They read
\begin{align}
    \langle \Phi_k \tilde{\Phi}_{k'} \rangle = \frac{1}{4h\sin^2(k/2) + r}N\delta_{k+k',0} &\hateq
    \tikz[baseline=-2.5pt]{
    \draw[Aactivity] (-0.7,0.0) -- (0.7,0.0)
    }
    \label{eq:prop1}\\
    \langle \Xi_k \bar{\Xi}_{k'} \rangle = - \langle \bar{\Xi}_k \Xi_{k'} \rangle = \frac{1 }{4h\sin^2(k/2) + r}N\delta_{k+k',0} &\hateq  
    \tikz[baseline=-2.5pt]{
    \draw[density] (-0.7,0.0) -- (0.7,0.0)
    }
    \label{eq:prop2}\\
    \langle \mathcal{J}_k \tilde{\mathcal{J}}_{k'}\rangle = N \delta_{k+k',0}&\hateq  
    \tikz[baseline=-2.5pt]{
    \draw[substrate] (-0.7,0.0) -- (0.7,0.0)
    }
    \label{eq:prop3}\\
    \langle \mathcal{J}_k \tilde{\Phi}_{k'}\rangle = \frac{h(-i\sin(k))}{4h\sin^2(k/2) + r}N\delta_{k+k',0} &\hateq  
    \tikz[baseline=-2.5pt]{
    \draw[Aactivity] (0.7,0.0) -- (0.0,0) ;
    \draw[substrate] (-0.7,0) -- (-0.0,0);
    \begin{scope}[rotate=180]
    \draw[Aactivity] (-0.15,-0.1) -- (-0.15,0.1);
    \end{scope}
    } ~,
    \label{eq:prop4}
\end{align}
where we have allowed for a mass $r$ for the purpose of infrared regularisation, which would enter the bilinear action $\mathcal{A}_0$ of Eq.~\eqref{eq:bilinear_action} via terms of the form $r \tilde{\phi}_i \phi_i$ and $r \bar{\xi}_i \xi_i$. All physical observables are evaluated in the limit of vanishing mass $r$ since the theory is massless as a matter of conservation of probability. The diagrammatic notation for the amputated interaction vertices is
\begin{alignat}{5}
    \gamma (\partial \tilde{\phi})^2 \phi^2 &\hateq
    \tikz[baseline=-2.5pt]{
    \draw[Aactivity] ($(40:0.4)+(0.0,0.0)$) -- (0.0,0.0) ;
    \draw[Aactivity] ($(-40:0.4)+(0.0,0.0)$) -- (0.0,0.0);
    \draw[Aactivity] ($(140:0.4)+(-0.0,0.0)$) -- (-0.0,0.0) ;
    \draw[Aactivity] ($(220:0.4)+(-0.0,0.0)$) -- (-0.0,0.0);
    \begin{scope}[rotate=-40]
    \draw[Aactivity] (-0.2,-0.1) -- (-0.2,0.1);
    \end{scope}
    \begin{scope}[rotate=+40]
    \draw[Aactivity] (-0.2,-0.1) -- (-0.2,0.1);
    \end{scope}
    }, \qquad
    \kappa \tilde{J}^2 \phi^2 &&\hateq
    \tikz[baseline=-2.5pt]{
    \draw[Aactivity] ($(40:0.4)+(0.0,0.0)$) -- (0.0,0.0) ;
    \draw[Aactivity] ($(-40:0.4)+(0.0,0.0)$) -- (0.0,0.0);
    \draw[substrate] ($(140:0.4)+(-0.0,0.0)$) -- (-0.0,0.0) ;
    \draw[substrate] ($(220:0.4)+(-0.0,0.0)$) -- (-0.0,0.0);
    }, \qquad
    \sigma \tilde{J}(\partial \tilde{\phi}) \phi^2 &&\hateq 
    \tikz[baseline=-2.5pt]{
    \draw[Aactivity] ($(40:0.4)+(0.0,0.0)$) -- (0.0,0.0) ;
    \draw[Aactivity] ($(-40:0.4)+(0.0,0.0)$) -- (0.0,0.0);
    \draw[substrate] ($(140:0.4)+(-0.0,0.0)$) -- (-0.0,0.0) ;
    \draw[Aactivity] ($(220:0.4)+(-0.0,0.0)$) -- (-0.0,0.0);
    \begin{scope}[rotate=+40]
    \draw[Aactivity] (-0.2,-0.1) -- (-0.2,0.1);
    \end{scope}
    }, \qquad
    \pi \tilde{J} (\partial \bar{\xi})\phi \xi &&\hateq 
    \tikz[baseline=-2.5pt]{
    \draw[Aactivity] ($(40:0.4)+(0.0,0.0)$) -- (0.0,0.0) ;
    \draw[density] ($(-40:0.4)+(0.0,0.0)$) -- (0.0,0.0);
    \draw[substrate] ($(140:0.4)+(-0.0,0.0)$) -- (-0.0,0.0) ;
    \draw[density] ($(220:0.4)+(-0.0,0.0)$) -- (-0.0,0.0);
    \begin{scope}[rotate=+40]
    \draw[Ddensity] (-0.2,-0.1) -- (-0.2,0.1);
    \end{scope}
    }, \qquad
    \chi (\partial \tilde{\phi})(\partial\bar{\xi})\phi \xi &&\hateq 
    \tikz[baseline=-2.5pt]{
    \draw[Aactivity] ($(40:0.4)+(0.0,0.0)$) -- (0.0,0.0) ;
    \draw[density] ($(-40:0.4)+(0.0,0.0)$) -- (0.0,0.0);
    \draw[Aactivity] ($(140:0.4)+(-0.0,0.0)$) -- (-0.0,0.0) ;
    \draw[density] ($(220:0.4)+(-0.0,0.0)$) -- (-0.0,0.0);
    \begin{scope}[rotate=+40]
    \draw[Ddensity] (-0.2,-0.1) -- (-0.2,0.1);
    \end{scope}
    \begin{scope}[rotate=-40]
    \draw[Aactivity] (-0.2,-0.1) -- (-0.2,0.1);
    \end{scope}
    } ~, \nonumber \\
    \gamma' (\partial \tilde{\phi})^2 \phi &\hateq
    \tikz[baseline=-2.5pt]{
    \draw[Aactivity] ($(0:0.4)+(0.0,0.0)$) -- (0.0,0.0) ;
    \draw[Aactivity] ($(140:0.4)+(-0.0,0.0)$) -- (-0.0,0.0) ;
    \draw[Aactivity] ($(220:0.4)+(-0.0,0.0)$) -- (-0.0,0.0);
    \begin{scope}[rotate=-40]
    \draw[Aactivity] (-0.2,-0.1) -- (-0.2,0.1);
    \end{scope}
    \begin{scope}[rotate=+40]
    \draw[Aactivity] (-0.2,-0.1) -- (-0.2,0.1);
    \end{scope}
    }, \qquad
    \kappa' \tilde{J}^2 \phi &&\hateq
    \tikz[baseline=-2.5pt]{
    \draw[Aactivity] ($(0:0.4)+(0.0,0.0)$) -- (0.0,0.0) ;
    \draw[substrate] ($(140:0.4)+(-0.0,0.0)$) -- (-0.0,0.0) ;
    \draw[substrate] ($(220:0.4)+(-0.0,0.0)$) -- (-0.0,0.0);
    }, \qquad
    \sigma' \tilde{J}(\partial \tilde{\phi}) \phi &&\hateq 
    \tikz[baseline=-2.5pt]{
    \draw[Aactivity] ($(0:0.4)+(0.0,0.0)$) -- (0.0,0.0) ;
    \draw[substrate] ($(140:0.4)+(-0.0,0.0)$) -- (-0.0,0.0) ;
    \draw[Aactivity] ($(220:0.4)+(-0.0,0.0)$) -- (-0.0,0.0);
    \begin{scope}[rotate=+40]
    \draw[Aactivity] (-0.2,-0.1) -- (-0.2,0.1);
    \end{scope}
    }, \qquad
    \pi' \tilde{J} (\partial \bar{\xi}) \xi &&\hateq 
    \tikz[baseline=-2.5pt]{
    \draw[density] ($(0:0.4)+(0.0,0.0)$) -- (0.0,0.0) ;
    \draw[substrate] ($(140:0.4)+(-0.0,0.0)$) -- (-0.0,0.0) ;
    \draw[density] ($(220:0.4)+(-0.0,0.0)$) -- (-0.0,0.0);
    \begin{scope}[rotate=+40]
    \draw[Ddensity] (-0.2,-0.1) -- (-0.2,0.1);
    \end{scope}
    }, \qquad
    \chi' (\partial \tilde{\phi})(\partial\bar{\xi}) \xi &&\hateq 
    \tikz[baseline=-2.5pt]{
    \draw[density] ($(0:0.4)+(0.0,0.0)$) -- (0.0,0.0) ;
    \draw[Aactivity] ($(140:0.4)+(-0.0,0.0)$) -- (-0.0,0.0) ;
    \draw[density] ($(220:0.4)+(-0.0,0.0)$) -- (-0.0,0.0);
    \begin{scope}[rotate=+40]
    \draw[Ddensity] (-0.2,-0.1) -- (-0.2,0.1);
    \end{scope}
    \begin{scope}[rotate=-40]
    \draw[Aactivity] (-0.2,-0.1) -- (-0.2,0.1);
    \end{scope}
    } ~, \nonumber \\
    \gamma'' (\partial \tilde{\phi})^2 &\hateq
    \tikz[baseline=-2.5pt]{
    \draw[Aactivity] ($(140:0.4)+(-0.0,0.0)$) -- (-0.0,0.0) ;
    \draw[Aactivity] ($(220:0.4)+(-0.0,0.0)$) -- (-0.0,0.0);
    \begin{scope}[rotate=-40]
    \draw[Aactivity] (-0.2,-0.1) -- (-0.2,0.1);
    \end{scope}
    \begin{scope}[rotate=+40]
    \draw[Aactivity] (-0.2,-0.1) -- (-0.2,0.1);
    \end{scope}
    }, \qquad\qquad
    \kappa'' \tilde{J}^2 &&\hateq
    \tikz[baseline=-2.5pt]{
    \draw[substrate] ($(140:0.4)+(-0.0,0.0)$) -- (-0.0,0.0) ;
    \draw[substrate] ($(220:0.4)+(-0.0,0.0)$) -- (-0.0,0.0);
    }, \qquad\qquad
    \sigma'' \tilde{J}(\partial \tilde{\phi})&&\hateq 
    \tikz[baseline=-2.5pt]{
    \draw[substrate] ($(140:0.4)+(-0.0,0.0)$) -- (-0.0,0.0) ;
    \draw[Aactivity] ($(220:0.4)+(-0.0,0.0)$) -- (-0.0,0.0);
    \begin{scope}[rotate=+40]
    \draw[Aactivity] (-0.2,-0.1) -- (-0.2,0.1);
    \end{scope}
    } \label{eq:vert_list}~.
\end{alignat}
with dashed propagators denoting spatial derivatives. At bare level, $\gamma = \kappa = 2\lambda$ and $\sigma = \pi = \chi = 4 \lambda$. Vertices appearing in the same column of the list \eqref{eq:vert_list} are generated from the same interaction term of Eq.~\eqref{eq:int_action} upon performing the shift $\phi_i \to \phi_i + N^{-1}$. Their coupling are thus related via $\gamma''/\gamma = \kappa''/\kappa = \sigma''/\sigma = N^{-2}$, $\gamma'/\gamma = \kappa'/\kappa=\sigma'/\sigma = 2 N^{-1}$ and $\pi'/\pi=\chi'/\chi = N^{-1}$. 

\subsection{Correlation functions in one dimension}
We are now ready to calculate the density and current correlation functions for the one-dimensional model. These are expressed as a power series in the small disorder strength $\lambda$, which we truncate to first order (tree level) to allow for direct comparison with the result obtained by expanding the analytical results by Derrida \cite{derrida_velocity_1983}, Eq.~\eqref{eq:st_st_flux1d}. This approximation is controlled in $d=1$ because we are interested in system of finite size $N$ and will be justified on relevance grounds in higher dimensions, where we will work in the continuum limit. 
The correlation function for the probability mass function at tree level is given by
\begin{align}
\label{eq:corr_fn_den}
    \langle \Phi_k \Phi_{k'} \rangle = 
    \tikz[baseline=-2.5pt]{
    \draw[Aactivity] (140:0.7) -- (0,0);
    \draw[Aactivity] (220:0.7) -- (0,0);
    \begin{scope}[rotate=-40]
    \draw[Aactivity] (-0.2,-0.1) -- (-0.2,0.1);
    \end{scope}
    \begin{scope}[rotate=+40]
    \draw[Aactivity] (-0.2,-0.1) -- (-0.2,0.1);
    \end{scope}
    } + \mathcal{O}(\lambda^2)
    = \frac{4 \lambda}{(2N h \sin(k/2))^2} N \delta_{k+k',0} + \mathcal{O}(\lambda^2)
\end{align}
for $k \neq 0$. For the current we have
\begin{align}
    \langle \mathcal{J}_k \mathcal{J}_{k'}\rangle = 
    \tikz[baseline=-2.5pt]{
    \draw[substrate] (140:0.7) -- (0,0);
    \draw[substrate] (220:0.7) -- (0,0);
    }
    +
    \tikz[baseline=-2.5pt]{
    \draw[substrate] (140:0.7) -- (0,0);
    \draw[Aactivity] (220:0.30) -- (0,0);
    \draw[substrate] ($(220:0.40)+(220:0.30)$) -- (220:0.30);
    \begin{scope}[rotate=+40]
    \draw[Aactivity] (-0.1,-0.1) -- (-0.1,0.1);
    \draw[Aactivity] (-0.2,-0.1) -- (-0.2,0.1);
    \end{scope}
    }
    +
    \tikz[baseline=-2.5pt]{
    \draw[Aactivity] (140:0.30) -- (0,0);
    \draw[substrate] ($(140:0.40)+(140:0.30)$) -- (140:0.30);
    \draw[Aactivity] (220:0.30) -- (0,0);
    \draw[substrate] ($(220:0.40)+(220:0.30)$) -- (220:0.30);
    \begin{scope}[rotate=+40]
    \draw[Aactivity] (-0.1,-0.1) -- (-0.1,0.1);
    \draw[Aactivity] (-0.2,-0.1) -- (-0.2,0.1);
    \end{scope}
    \begin{scope}[rotate=-40]
    \draw[Aactivity] (-0.1,-0.1) -- (-0.1,0.1);
    \draw[Aactivity] (-0.2,-0.1) -- (-0.2,0.1);
    \end{scope}
    }+ \mathcal{O}(\lambda^2)
    = \frac{4 \lambda}{N^2} N \delta_{k+k',0} \delta_{k',0} + \mathcal{O}(\lambda^2)~, \label{eq:curr_cov_diagrs}
\end{align}
with Kronecker delta $\delta_{k',0}$ reflecting the constraint that $J_i = J_j = J$ is homogeneous. For $k \neq 0$, the three diagrams contributing to the right-hand side of Eq.~\eqref{eq:curr_cov_diagrs} only differ in their symmetry factors and overall sign, which produces the desired cancellation. When $k=0$, only the first diagram, which does not involve spatial derivatives, contributes and the cancellation does not occur.
Since $\mathcal{J}_{k=0} = N J$, we thus have that $\langle J^2 \rangle = N^{-2} \langle \mathcal{J}_0 \mathcal{J}_0 \rangle = 4 \lambda N^{-3}$, recovering the leading order contribution obtained by expanding Derrida's \cite{derrida_velocity_1983} exact result, Eq.~\eqref{eq:st_st_flux1d}, with the antisymmetric condition $\zeta_{i,i+1} - \zeta_{i,i-1}= 2 \zeta_i$ , 
\begin{align}
    \langle J^2 \rangle = \frac{1}{N^4} \left\langle \left( \sum_{n=1}^N \zeta_{n,n+1} - \zeta_{n+1,n} \right) \left( \sum_{m=1}^N \zeta_{m,m+1} - \zeta_{m+1,m} \right)\right\rangle 
    = \frac{4}{N^4} \sum_{n=1}^N \sum_{m=1}^N \langle \zeta_n \zeta_m \rangle = \frac{4 \lambda}{N^3} ~.
\end{align}

\subsection{Higher dimensions}
The derivation presented above can be straighforwardly generalised to periodic lattices of arbitrary dimension with sites $\bm{i}= (i_1,...,i_d) \in \{1,...,N\}^d$. We further need to modify the steady-state condition, Eq.~\eqref{ss_def}, using higher dimensional extensions of Eqs.~\eqref{eq:discr_1d} and \eqref{eq:discr_2d}, $\bm{\nabla}_d = (\partial_{i_1},...,\partial_{i_d})$ and $\Delta_d = \partial^2_{i_1}+...+\partial^2_{i_d}$, so that
\begin{equation}
0 = \partial_t \phi_{\bm{i}} = h \Delta_d \phi_{\bm{i}} - 2 \bm{\nabla}_d . (\bm{\zeta}_{\bm{i}} \phi_{\bm{i}})
\end{equation}
where $\bm{\zeta}_{\bm{i}} = (\zeta^{(1)}_{\bm{i}},...,\zeta^{(d)}_{\bm{i}})\in \mathbb{R}^d$ is a $d$-dimensional noise with correlator
\begin{equation}
    \langle \zeta^{(a)}_{\bm{k}} \zeta^{(b)}_{\bm{k}'} \rangle = \lambda \delta_{ab} N^d \delta_{\bm{k}+\bm{k}',0}~.
\end{equation}
Here, $\zeta^{(a)}_{\bm{i}}$ denotes the noise affecting the edges of node $\bm{i}$ along the $a$th dimension of the lattice and $\zeta^{(a)}_{\bm{k}}$ its Fourier transform, according to the convention
\begin{equation}
   \zeta^{(a)}_{\bm{j}} = \frac{1}{N^d} \sum_{\bm{k}} e^{i \bm{k} \cdot \bm{j}} \zeta^{(a)}_{\bm{k}} \quad \text{and}  \quad \zeta^{(a)}_{\bm{k}} = \sum_{\bm{j}} e^{- i \bm{k} \cdot \bm{j}} \zeta^{(a)}_{\bm{j}} ~,
\end{equation}
with $\bm{k} = 2 \pi \bm{n}/N$ and $\bm{n} \in \{1,...,N\}^d$.
Correspondingly Eq.~\eqref{eq:curr_mean_def} is modified to define a vector lattice field $\bm{J}_{\bm{i}} \in \mathbb{R}^d$. Writing $\bm{J}_{\bm{i}} = (J_{\bm{i}}^{(1)},...,J_{\bm{i}}^{(d)})$, 
\begin{equation}
    J_{\bm{i}}^{(a)} = -h \partial_{i_a} \phi_{\bm{i}} + \frac{1}{2} \partial_{i_a}^2 (\zeta^{(a)}_{\bm{i}} \phi_{\bm{i}}) + 2 \zeta^{(a)}_{\bm{i}} \phi_{\bm{i}}
    \label{eq:id2_generald}~,
\end{equation}
where $J_{\bm{i}}^{(a)}$ is once again the symmetric local average at site $\bm{i}$ of the probability current along the $a$th dimension of the lattice. 
The functional relationship between each component of the current vector field and the other fields, Eq.~\eqref{eq:id2_generald}, is then imposed by means of $d$ resolutions of the identity of the form \eqref{identity_2}. In the case of uncorrelated noise, the derivation of the action follows precisely the same lines as above and generates $d$ copies of the type of terms we have already seen, each with spatial derivatives taken with respect to a different dimension. Explicitly,
\begin{equation}
\label{eq:bil_act_d_dim}
    \mathcal{A}_0 = \sum_{\bm{i}} -h \tilde{\phi}_{\bm{i}} \Delta_d \phi_{\bm{i}} -h \bar{\xi}_{\bm{i}} \Delta_d \xi_{\bm{i}} + \tilde{\bm{J}}_{\bm{i}} \cdot \bm{J}_{\bm{i}} + h \tilde{\bm{J}}_{\bm{i}} \cdot \bm{\nabla}_d\phi_{\bm{i}}
\end{equation}
{\it cf}.\ Eq.~\eqref{eq:bilinear_action} and
\begin{align}
\label{eq:int_act_d_dim}
    \mathcal{A}_{\rm int}
    &=  
    - 2\lambda \sum_{\bm{i}} [(\bm{\nabla}_d \tilde{\phi}_{\bm{i}})^2 \phi^2_{\bm{i}} + \tilde{\bm{J}}^2_{\bm{i}} \phi^2_{\bm{i}} + 2 \bm{\tilde{J}}_{\bm{i}} \cdot (\bm{\nabla}_d \tilde{\phi}_{\bm{i}}) \phi_{\bm{i}}^2 + 2 \bm{\tilde{J}}_{\bm{i}} \cdot (\bm{\nabla}_d \bar{\xi}_{\bm{i}}) \phi_{\bm{i}} \xi_{\bm{i}} + 2 (\bm{\nabla}_d \tilde{\phi}_{\bm{i}}) \cdot (\bm{\nabla}_d \bar{\xi}_{\bm{i}}) \phi_{\bm{i}} \xi_{\bm{i}}]
    + \rm{h.o.t.} ~,
\end{align}
{\it cf}.\ Eq.~\eqref{eq:int_action}. Since vertices 
arising from Eq.~\eqref{eq:int_act_d_dim} involve vector fields, the intepretation of diagrams
needs to be modified slightly. For example, the mixed propagators in Fourier representation becomes
\begin{equation}
    \langle \mathcal{J}^{(a)}_{\bm{k}} \tilde{\Phi}_{\bm{k'}}\rangle = \frac{N^d h(-i \sin(k_a))}{4h \sum_b \sin^2(k_b/2) + r} \delta_{\bm{k}+\bm{k'},0}\hateq
    \tikz[baseline=-2.5pt]{
    \draw[Aactivity] (0.7,0.0) -- (0.0,0) ;
    \draw[substrate] (-0.7,0) -- (-0.0,0);
    \begin{scope}[rotate=180]
    \draw[Aactivity] (-0.15,-0.1) -- (-0.15,0.1);
    \end{scope}
    } ~.
\end{equation}
Similarly,
\begin{equation}
    (\tilde{\bm{J}}_{\bm{i}})^2 \phi_{\bm{i}}^2 \hateq
    \tikz[baseline=-2.5pt]{
    \draw[substrate] (140:0.4) -- (0,0);
    \draw[substrate] (220:0.4) -- (0,0);
    \draw[Aactivity] (40:0.4) -- (0,0);
    \draw[Aactivity] (-40:0.4) -- (0,0);
    } , \quad
    \tilde{\bm{J}}_{\bm{i}} \cdot (\bm{\nabla}_d \tilde{\phi}_{\bm{i}}) \phi_{\bm{i}}^2 \hateq 
    \tikz[baseline=-2.5pt]{
    \draw[substrate] (140:0.4) -- (0,0);
    \draw[Aactivity] (220:0.4) -- (0,0);
    \draw[Aactivity] (40:0.4) -- (0,0);
    \draw[Aactivity] (-40:0.4) -- (0,0);
    \begin{scope}[rotate=+40]
    \draw[Aactivity] (-0.2,-0.1) -- (-0.2,0.1);
    \end{scope}
    } , \quad
    (\bm{\nabla}_d \tilde{\phi}_{\bm{i}})^2 \phi_{\bm{i}}^2 \hateq 
    \tikz[baseline=-2.5pt]{
    \draw[Aactivity] (140:0.4) -- (0,0);
    \draw[Aactivity] (220:0.4) -- (0,0);
    \draw[Aactivity] (40:0.4) -- (0,0);
    \draw[Aactivity] (-40:0.4) -- (0,0);
    \begin{scope}[rotate=-40]
    \draw[Aactivity] (-0.2,-0.1) -- (-0.2,0.1);
    \end{scope}
    \begin{scope}[rotate=+40]
    \draw[Aactivity] (-0.2,-0.1) -- (-0.2,0.1);
    \end{scope}
    }, \quad
    \tilde{\bm{J}}_{\bm{i}} \cdot (\bm{\nabla}_d \bar{\xi}_{\bm{i}}) \phi_{\bm{i}} \xi_{\bm{i}} \hateq 
    \tikz[baseline=-2.5pt]{
    \draw[substrate] (140:0.4) -- (0,0);
    \draw[density] (220:0.4) -- (0,0);
    \draw[Aactivity] (40:0.4) -- (0,0);
    \draw[density] (-40:0.4) -- (0,0);
    \begin{scope}[rotate=+40]
    \draw[Ddensity] (-0.2,-0.1) -- (-0.2,0.1);
    \end{scope}
    }, \quad
    (\bm{\nabla}_d \tilde{\phi}_{\bm{i}}) \cdot (\bm{\nabla}_d \bar{\xi}_{\bm{i}}) \phi_{\bm{i}} \xi_{\bm{i}} \hateq 
    \tikz[baseline=-2.5pt]{
    \draw[Aactivity] (140:0.4) -- (0,0);
    \draw[density] (220:0.4) -- (0,0);
    \draw[Aactivity] (40:0.4) -- (0,0);
    \draw[density] (-40:0.4) -- (0,0);
    \begin{scope}[rotate=-40]
    \draw[Aactivity] (-0.2,-0.1) -- (-0.2,0.1);
    \end{scope}
    \begin{scope}[rotate=+40]
    \draw[Ddensity] (-0.2,-0.1) -- (-0.2,0.1);
    \end{scope}
    } ~.
\end{equation}
Using the field theory in $d>1$ to characterise correlations is most covenient in the continuum limit. We thus take the limit $N \to \infty$ at fixed volume $V= (N\ell)^d $, with $\ell$ the dimensionful lattice spacing. Based on Eqs.~\eqref{eq:discr_1d} and \eqref{eq:discr_2d} we define the differential operators in the continuum $\bm{\nabla}_d = \ell \bm{\nabla}$ and $\Delta_d = \ell^2 \Delta$, together with the fields $\phi$ and $\tilde{\phi}$ now being defined for all $\bm{x} \in (0,N\ell]^d$, such that
\begin{equation}
    \phi(\bm{i}\ell) = \ell^{-d} \phi_{\bm{i}} , \quad \tilde{\phi}(\bm{i}\ell) = \ell^\beta \tilde{\phi}_{\bm{i}} ~,
\end{equation}
where the dimension of $\phi(\bm{x})$ is pre-determined by the shift performed above, which in the continuum reads $\phi(\bm{x}) \to V^{-1} + \phi(\bm{x})$, while the exponent $\beta$ is undetermined for the time being.
Converting sums over lattice sites into integrals over space according to
\begin{equation}
    \ell^d \sum_{\bm{i}} \equiv \int_V d^dx
\end{equation}
and demanding that the action of Eqs.~\eqref{eq:bil_act_d_dim} and \eqref{eq:int_act_d_dim} is dimensionless term by term, defines the remaining fields
\begin{equation}
    \bm{J}(\bm{i}\ell) = \ell^{-d+1-\beta} \bm{J}_{\bm{i}}, \quad
    \tilde{\bm{J}}(\bm{i}\ell) = \ell^{-1+\beta} \tilde{\bm{J}}_{\bm{i}} , \quad 
    \bar{\xi}(\bm{i}\ell) \xi(\bm{i}\ell) = \ell^{\beta-d} \bar{\xi}_{\bm{i}} \xi_{\bm{i}} 
\end{equation}
and the dimensionful couplings of the continuum theory $D = h \ell^{2-\beta}$ and $\tilde{\lambda} = \ell^{d+2-2\beta} \lambda$. The exponent $\beta$ is fixed by imposing that the diffusion constant $D$ remains finite in the limit $\ell \to 0$, whence $\beta = 2$. Since now $\tilde{\lambda} = \lambda \ell^{d-2}$, we identify $d_c = 2$ as the upper critical dimension of the theory. For $d>2$ all interactions become irrelevant and we expect the large scale behaviour of the theory to be well described by the tree-level diagrammatics, thus justifying the truncation of higher order terms. Precisely at $d=2$, renormalisation group theory \cite{zinn2021quantum} predicts logarithmic corrections to tree-level scaling but these were not observed conclusively in our numerical investigation. 

In the continuum we thus have the action $\mathcal{A}_0 + \mathcal{A}_{\rm int}$ with
\begin{align}
    \mathcal{A}_0 &= - \int_V d^dx \ D \tilde{\phi}(\bm{x}) \Delta \phi(\bm{x}) + D \bar{\xi}(\bm{x}) \Delta \xi(\bm{x}) - \tilde{\bm{J}}(\bm{x}) \cdot \bm{J}(\bm{x}) - D \tilde{\bm{J}}(\bm{x}) \cdot \bm{\nabla} \phi(\bm{x}) \\
    \mathcal{A}_{\rm{int}} &=  - 2\tilde{\lambda} \int_V d^dx \ [(\bm{\nabla} \tilde{\phi}(\bm{x}))^2 (\phi(\bm{x}) + V^{-1})^2 + \tilde{\bm{J}}^2(\bm{x}) (\phi(\bm{x}) + V^{-1})^2 + 2 \bm{\tilde{J}}(\bm{x}) \cdot (\bm{\nabla} \tilde{\phi}(\bm{x})) (\phi(\bm{x}) + V^{-1})^2 \nonumber \\
    &\quad\quad\quad\quad + 2 \bm{\tilde{J}}(\bm{x}) \cdot (\bm{\nabla} \bar{\xi}(\bm{x})) (\phi(\bm{x}) + V^{-1}) \xi(\bm{x}) + 2 (\bm{\nabla} \tilde{\phi}(\bm{x})) \cdot (\bm{\nabla} \bar{\xi}(\bm{x})) (\phi(\bm{x}) + V^{-1}) \xi(\bm{x})]
    + \rm{h.o.t.} ~.
\end{align}
Absorbing the shift $V^{-1}$ of $\phi$ into new couplings as done using dashed variables in Eq.~\eqref{eq:vert_list} means that these differ now from each other in their engineering dimension, for example $[\gamma] = [V \gamma'] = [V^2 \gamma'']$.
Maintaining a finite volume $V$ is important for the validity of the present theory, but it also means that Fourier transforming it by mean of Eq.~\eqref{eq:conv_ft} results in sums over suitable modes and $\delta$-functions of the form
\begin{equation}
    V \delta_{\bm{k},0} = \int_V d^dx \ e^{-i \bm{k} \cdot \bm{x}}, \quad \text{whence e.g.} \quad \Phi(\bm{k}) = \int_V d^dx \ e^{-i \bm{k}\cdot \bm{x}} \phi(\bm{x}) ~.
\end{equation}
We may occasionally approximate sums over momenta by integrals. To make the notation more suggestive of continuous $\bm{k} \in \mathbb{R}^d$ we write the Fourier transformed fields as $\Phi(\bm{k})$, $\bm{\mathcal{J}}(\bm{k})$, $\xi(\bm{k})$ etc.

The continuum propagators are structurally identical to their discrete counterpart and read
\begin{align}
    \langle \Phi(\bm{k}) \tilde{\Phi}(\bm{k}') \rangle = \frac{V \delta_{\bm{k}+\bm{k'},0}}{D|\bm{k}|^2 + r} &\hateq
    \tikz[baseline=-2.5pt]{
    \draw[Aactivity] (-0.7,0.0) -- (0.7,0.0)
    }
    \label{eq:prop1_cont}\\
    \langle \Xi(\bm{k}) \bar{\Xi}(\bm{k}') \rangle = - \langle \bar{\Xi}(\bm{k}) \Xi(\bm{k}') \rangle = \frac{V \delta_{\bm{k}+\bm{k'},0} }{D|\bm{k}|^2 + r} &\hateq  
    \tikz[baseline=-2.5pt]{
    \draw[density] (-0.7,0.0) -- (0.7,0.0)
    }
    \label{eq:prop2_cont}\\
    \langle {\mathcal{J}}^{(n)}(\bm{k}) \tilde{{\mathcal{J}}}^{(m)}(\bm{k}')\rangle = \delta_{nm} V \delta_{\bm{k}+\bm{k'},0}&\hateq  
    \tikz[baseline=-2.5pt]{
    \draw[substrate] (-0.7,0.0) -- (0.7,0.0)
    }
    \label{eq:prop3_cont}\\
    \langle \bm{\mathcal{J}}(\bm{k}) \tilde{\Phi}(\bm{k}')\rangle = \frac{ D(-i\bm{k})}{D|\bm{k}|^2 + r} V\delta_{\bm{k}+\bm{k'},0} &\hateq  
    \tikz[baseline=-2.5pt]{
    \draw[Aactivity] (0.7,0.0) -- (0.0,0) ;
    \draw[substrate] (-0.7,0) -- (-0.0,0);
    \begin{scope}[rotate=180]
    \draw[Aactivity] (-0.15,-0.1) -- (-0.15,0.1);
    \end{scope}
    } ~.
    \label{eq:prop4_cont}
\end{align}

\subsection{Correlated noise}
Generalising the formalism to allow for correlations in the transition rates is relatively straightforward. What changes is the form of the probability functional for the noise, originally Eq.~\eqref{gaussian_prob}, which should now include a non-trivial dependence on the power spectrum. The power spectrum $Q(\bm{k})$ characterises the noise correlation function for the different components of the disorder in Fourier space as
\begin{equation}
\label{eq:noise_corr_mom}
    \langle \zeta^{(n)}(\bm{k}) \zeta^{(m)}(\bm{p}) \rangle = Q(\bm{k}) V \delta_{\bm{k}+\bm{p},0} \delta_{n,m} ~.
\end{equation}
The case of uncorrelated noise covered above corresponds to $Q(\bm{k}) = \tilde{\lambda}$, i.e.\ to the case where $Q(\bm{k})$ is independent of $\bm{k}$. The noise probability functional corresponding to Eq.~\eqref{eq:noise_corr_mom} reads
\begin{equation}
\label{eq:corr_noise_weight}
    P[\bm{\zeta}] \propto \exp \left[ - \frac{1}{2V}\sum_{\bm{k}} \frac{ \bm{\zeta}(\bm{k}) \cdot \bm{\zeta}(-\bm{k})}{Q(\bm{k})} \right]
\end{equation}
where the sample space of $\bm{\zeta}(\bm{k})$ is constrained to $\bm{\zeta}(\bm{k}) = \bm{\zeta}^*(-\bm{k})$ as to maintain $\bm{\zeta}(\bm{x})$, the inverse Fourier transform of $\bm{\zeta}(\bm{k})$, being real. The role of $Q(\bm{k})$ is to penalise spatial fluctuations of $\bm{\zeta}(\bm{k})$ depending on its wavenumber.
Since our setup is rotationally symmetric we assume the general algebraic form of the correlator to be $Q(k) \sim \tilde{\lambda} k^{-\eta}$ for small $k$, where $k = |\bm{k}|$.
For $\eta > 0$, the right hand side of Eq.~\eqref{eq:corr_noise_weight} is regularised by excluding the homogeneous mode, $\bm{k}=0$, from the summation.
The derivation of the field theory with non-trivial correlator proceeds as outline above up until Eq.~\eqref{noise_integr}, which has now a slightly different form,
\begin{align}
    \int \mathcal{D}\zeta \ &{\rm exp}\left( \frac{1}{V}\sum_{\bm{k}} - \frac{k^\eta}{2 \tilde{\lambda}} \bm{\zeta}(\bm{k}) \cdot \bm{\zeta}(-\bm{k}) + \bm{\zeta}(\bm{k}) \cdot \bm{F}(-\bm{k}) \right) \\
    \propto \ &{\rm exp}\left( \frac{\tilde{\lambda}}{2V} \sum_{\bm{k}}  k^{-\eta} \bm{F}(\bm{k}) \cdot \bm{F}(-\bm{k}) \right)~,
\end{align}
where we have introduced the short-hand $\bm{F}(\bm{k})$ for the Fourier transform of the real expression
\begin{equation}
\bm{F}(\bm{x}) = 2(\bm{\nabla} \tilde{\phi}(\bm{x}))\phi(\bm{x}) + 2\tilde{\bm{J}}(\bm{x})\phi(\bm{x})  + 2(\bm{\nabla} \bar{\xi}(\bm{x}))\xi(\bm{x}) + \rm{h.o.t}~. 
\end{equation}
We thus find that the generalisation to correlated noise characterised by a power spectrum $Q(k) = \tilde{\lambda} k^{-\eta} $ amounts to augmenting the interaction terms (in Fourier space) of the original theory by a factor $k^{-\eta}$. The Fourier representation, originally Eq.~\eqref{eq:exmpl_ft_int}, of the interaction vertices to be used in our subsequent calculation is upgraded to the its continuum form for general $\eta$ according to
\begin{align}
    \tikz[baseline=-2.5pt]{
    \draw[Aactivity] ($(140:0.4)+(-0.0,0.0)$) -- (-0.0,0.0) ;
    \draw[Aactivity] ($(220:0.4)+(-0.0,0.0)$) -- (-0.0,0.0);
    \begin{scope}[rotate=-40]
    \draw[Aactivity] (-0.2,-0.1) -- (-0.2,0.1);
    \end{scope}
    \begin{scope}[rotate=+40]
    \draw[Aactivity] (-0.2,-0.1) -- (-0.2,0.1);
    \end{scope}
    }
    &\hateq (\gamma''/V)  k^{-\eta} (-i \bm{k}) \cdot (i \bm{k}) \tilde{\Phi}(\bm{k}) \tilde{\Phi}(-\bm{k}) 
    \\
    \tikz[baseline=-2.5pt]{
    \draw[substrate] ($(140:0.4)+(-0.0,0.0)$) -- (-0.0,0.0) ;
    \draw[substrate] ($(220:0.4)+(-0.0,0.0)$) -- (-0.0,0.0);
    }
    &\hateq (\kappa''/V) k^{-\eta} \tilde{\bm{\mathcal{J}}}(\bm{k}) \cdot \tilde{\bm{\mathcal{J}}}(-\bm{k}) 
    \\
    \tikz[baseline=-2.5pt]{
    \draw[substrate] ($(140:0.4)+(-0.0,0.0)$) -- (-0.0,0.0) ;
    \draw[Aactivity] ($(220:0.4)+(-0.0,0.0)$) -- (-0.0,0.0);
    \begin{scope}[rotate=+40]
    \draw[Aactivity] (-0.2,-0.1) -- (-0.2,0.1);
    \end{scope}
    }
    &\hateq (\sigma''/V) k^{-\eta} (-i \bm{k}) \cdot \tilde{\bm{\mathcal{J}}}(\bm{k}) \tilde{\Phi}(-\bm{k})
    ~,
\end{align}
with $\gamma'' = \kappa''= 2\tilde{\lambda}/V^2$ and $\sigma''=4\tilde{\lambda}/V^2$.
The effect of this modification on the spectral density of the currents becomes apparent already at tree level, where we now obtain, for $\bm{k} \neq 0$, 
\begin{align}
    \langle \mathcal{J}^{(n)}(\bm{k}) \mathcal{J}^{(m)}(\bm{k}') \rangle &= 
    \tikz[baseline=-2.5pt]{
    \draw[substrate] (140:0.7) -- (0,0);
    \draw[substrate] (220:0.7) -- (0,0);
    } 
    +
    \tikz[baseline=-2.5pt]{
    \draw[substrate] (140:0.7) -- (0,0);
    \draw[Aactivity] (220:0.30) -- (0,0);
    \draw[substrate] ($(220:0.40)+(220:0.30)$) -- (220:0.30);
    \begin{scope}[rotate=+40]
    \draw[Aactivity] (-0.1,-0.1) -- (-0.1,0.1);
    \draw[Aactivity] (-0.2,-0.1) -- (-0.2,0.1);
    \end{scope}
    }
    +
    \tikz[baseline=-2.5pt]{
    \draw[Aactivity] (140:0.30) -- (0,0);
    \draw[substrate] ($(140:0.40)+(140:0.30)$) -- (140:0.30);
    \draw[Aactivity] (220:0.30) -- (0,0);
    \draw[substrate] ($(220:0.40)+(220:0.30)$) -- (220:0.30);
    \begin{scope}[rotate=+40]
    \draw[Aactivity] (-0.1,-0.1) -- (-0.1,0.1);
    \draw[Aactivity] (-0.2,-0.1) -- (-0.2,0.1);
    \end{scope}
    \begin{scope}[rotate=-40]
    \draw[Aactivity] (-0.1,-0.1) -- (-0.1,0.1);
    \draw[Aactivity] (-0.2,-0.1) -- (-0.2,0.1);
    \end{scope}
    } + \mathcal{O}(\tilde{\lambda}^2/V^2)
    \\
    &= \left(\frac{4 \tilde{\lambda}}{V} \delta_{n,m} - \frac{8 \tilde{\lambda}}{V} \frac{k_n k_m}{|\bm{k}|^2} + \frac{4 \tilde{\lambda}}{V} \sum_{\ell=1}^d \frac{k_n k_m k_\ell^2}{|\bm{k}|^4} \right) |\bm{k}|^{-\eta} \delta_{\bm{k} + \bm{k}',0} + \mathcal{O}(\tilde{\lambda}^2/V^2) \\
    &=  \frac{4 \tilde{\lambda}}{V}  \left( \delta_{n,m} - \frac{k_n k_m}{|\bm{k}|^2} \right) |\bm{k}|^{-\eta} \delta_{\bm{k} + \bm{k}',0} + \mathcal{O}(\tilde{\lambda}^2/V^2)~.
    \label{eq:final_corr_fn_app_der}
\end{align}
This is the result we invoke in the main text, \Eqref{eq:sp_dens_j}. The special case of uncorrelated rates is recovered at $\eta = 0$. The momentum-dependent factor appearing in brackets in Eq.~\eqref{eq:final_corr_fn_app_der} is the fingerprint of a solenoidal vector field \cite{forster_large-distance_1977,monin1975}. For $\bm{k}=0$ and $\eta \leq 0$, the second and third diagrams vanish and we instead find
\begin{equation}
    \langle \mathcal{J}^{(n)}(0) \mathcal{J}^{(m)}(\bm{k}') \rangle = \frac{4 \tilde{\lambda}}{V^2} |\bm{k'}|^{-\eta} \delta_{n,m} V \delta_{\bm{k}',0}~,
\end{equation}
consistently with Eq.~\eqref{eq:curr_cov_diagrs}.

\section{Large $L$ asymptotic form of  Eq.~(18)\label{A:asym_lograt}} 
We start from \Eqref{eq:ave_simeso} for the expectation of the entropy production per mesostate with linear block dimensions $L$, where $\{\alpha,\beta\}$ is any pair of neighbouring mesostates. Without loss of generality we assume that the interface is a $(d-1)$-dimensional hypercube of constant $x$ coordinate and denote $\{\bm{i}_\alpha\}$ the set of ``boundary'' states of $\alpha$ connected to $\beta$ by a single edge. Denoting $\bm{e} = (1,0,\cdots,0)$ the $d$-dimensional unit displacement vector along the $x$ axis, let $\{\bm{i}_\alpha+\bm{e}\}$ correspond to the set of ``boundary'' states of $\beta$ connected to $\alpha$ by a single edge. Inspecting the logarithmic factor on the right-hand side of Eq.~\eqref{eq:ave_simeso},
\begin{align}
\ln \frac{j_{\alpha \beta}{(L)}}{j_{\beta \alpha}{(L)}} &= \ln \frac{\sum_{\{ \bm{i}_\alpha \}} h \bar{\pi} + h \delta \pi_{\bm{i}} + \pi_{\bm{i}} \zeta_{\bm{i},\bm{i}+\bm{e}}  }{   \sum_{\{ \bm{i}_\alpha \}} h \bar{\pi} + h \delta \pi_{\bm{i} + \bm{e}} + \pi_{\bm{i}+\bm{e}} \zeta_{\bm{i}+\bm{e},\bm{i}}} \\
&= \ln \left( 1 + \frac{\sum_{\{ \bm{i}_\alpha \}} h \delta \pi_{\bm{i}} + \pi_{\bm{i}} \zeta_{\bm{i},\bm{i}+\bm{e}}}{h \bar{\pi} L^{d-1} } \right) - \ln \left( 1 + \frac{\sum_{\{ \bm{i}_\alpha \}} h \delta \pi_{\bm{i} + \bm{e}} + \pi_{\bm{i}+\bm{e}} \zeta_{\bm{i}+\bm{e},\bm{i}}}{h \bar{\pi} L^{d-1} } \right) \label{eq:expanded_log}
\end{align}
where we have expanded the steady-state probability about its mean, $\pi_{\bm{i}} = \bar{\pi} + \delta \pi_{\bm{i}}$, with $\bar{\pi} = N^{-d}$. Each logarithm in the expression above contains a sum over $L^{d-1}$ zero-mean random variables $h \delta \pi_{\bm{i}} + \pi_{\bm{i}} \zeta_{\bm{i},\bm{i}+\bm{e}}$ and $h \delta \pi_{\bm{i} + \bm{e}} + \pi_{\bm{i}+\bm{e}} \zeta_{\bm{i}+\bm{e},\bm{i}}$, respectively. 
While the expected value of each sum vanishes by linearity, the typical magnitude can be estimated by computing its standard deviation. We thus write down the variance of the sum
\begin{align} \label{eq:var_sum_expand}
    \frac{1}{(h \bar{\pi} L^{d-1})^2 } \left\langle \left( \sum_{\{ \bm{i}_\alpha \}} h \delta \pi_{\bm{i} + \bm{e}} + \pi_{\bm{i}+\bm{e}} \zeta_{\bm{i}+\bm{e},\bm{i}} \right)^2 \right\rangle &= \frac{1}{(h \bar{\pi} L^{d-1})^2 }\sum_{\bm{i}_\alpha} \left\langle \left( h \delta \pi_{\bm{i} + \bm{e}} + \pi_{\bm{i}+\bm{e}} \zeta_{\bm{i}+\bm{e},\bm{i}} \right)^2 \right\rangle \nonumber\\
    &+ \frac{1}{(h \bar{\pi} L^{d-1})^2 }\sum_{\{\bm{i}_\alpha \neq {\bm{i}'}_\alpha\}} \left\langle \left( h \delta \pi_{\bm{i} + \bm{e}} + \pi_{\bm{i}+\bm{e}} \zeta_{\bm{i}+\bm{e},\bm{i}} \right) \left( h \delta \pi_{\bm{i}' + \bm{e}} + \pi_{\bm{i}'+\bm{e}} \zeta_{\bm{i}'+\bm{e},\bm{i}'} \right) \right\rangle ~.
\end{align}
The first contribution on the right hand side of Eq.~\eqref{eq:var_sum_expand}, associated with the sum of variances, scales like $L^{-(d-1)}$ due to statistical homogeneity and can thus be taken to be small compared to the factor $1$ appearing in the logarithm \Eqref{eq:expanded_log} in the large $L$ regime. The second contribution is associated with cross correlations among the summands. It involves $L^{d-1}(L^{d-1} - 1)$ terms so it is at most of order $L^0$. However, this scaling only applies to infinite-range correlations and the more relevant decay of density as well as noise correlations over large distances will in general produce a scaling $L^{\nu}$ with $\nu < 0$. The second contributions is thus also asymptotically small compared to the factor $1$ appearing in the logarithm. For sufficiently large $L$ we can therefore expand \Eqref{eq:expanded_log} to leading order and obtain
\begin{align}
    \ln \frac{j_{\alpha \beta}{(L)}}{j_{\beta \alpha}{(L)}} &\simeq \frac{1}{h \bar{\pi} L^{d-1} }  \sum_{\{ \bm{i}_\alpha \}} h \delta \pi_{\bm{i}} + \pi_{\bm{i}} \zeta_{\bm{i},\bm{i}+\bm{e}} - h \delta \pi_{\bm{i} + \bm{e}} - \pi_{\bm{i}+\bm{e}} \zeta_{\bm{i}+\bm{e},\bm{i}} \nonumber \\
    &= \frac{1}{h \bar{\pi} L^{d-1} } ( j_{\alpha \beta}{(L)} - j_{\beta \alpha}{(L)} ) ~,
\end{align}
whence
\begin{equation}
    \left\langle (j_{\alpha \beta}{(L)} - j_{\beta \alpha}{(L)}) \ln \frac{j_{\alpha \beta}{(L)}}{j_{\beta \alpha}{(L)}} \right\rangle \simeq \frac{N^d}{h L^{d-1} } \langle ( j_{\alpha \beta}{(L)} - j_{\beta \alpha}{(L)} )^2 \rangle ~.
\end{equation}

\section{Asymptotic scaling of integrated currents from correlation function \label{A:hyperu_scaling}}
\begin{figure}
    \centering
    \includegraphics[scale=0.4]{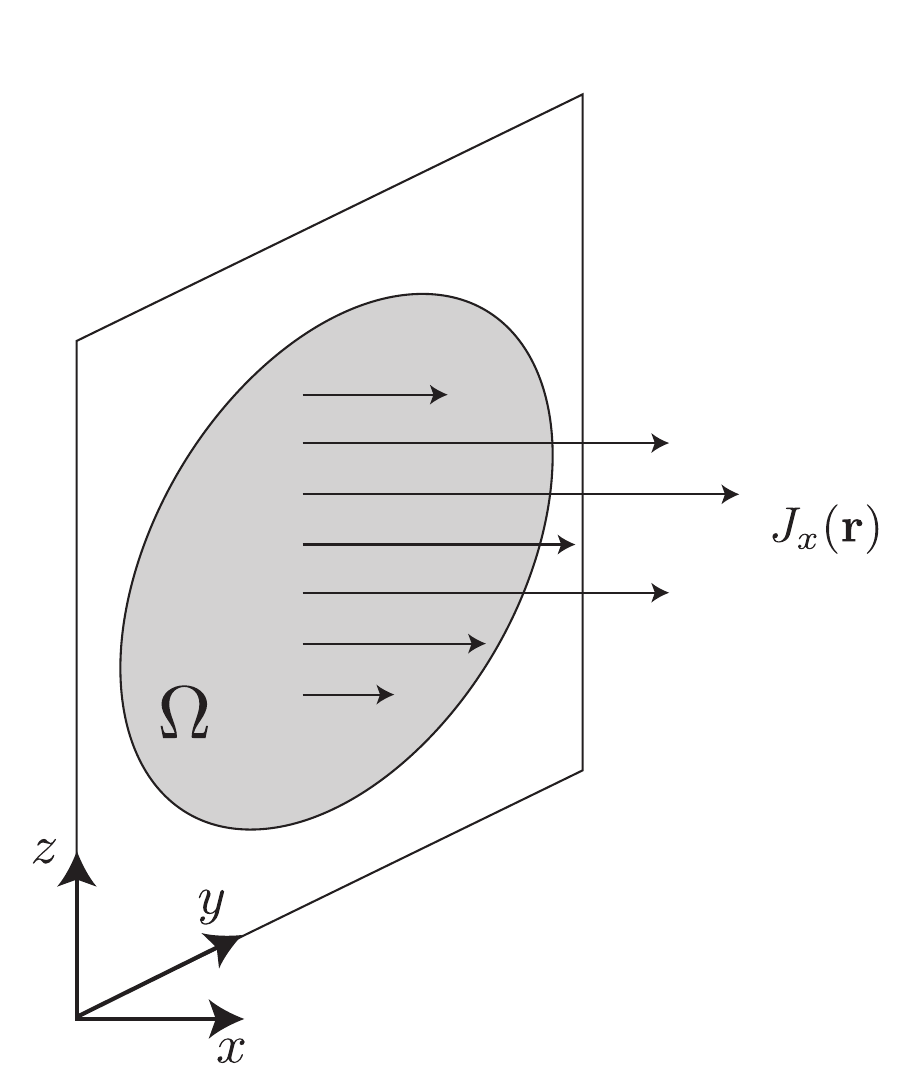}
    \caption{Schematic illustration of the setup used for the calculation of the statistics of the integrated current across a mesostate interface $\Omega$, which we assume to be a $d-1$ dimensional hypersurface of constant $x=0$ coordinate embedded in $d$ dimensional space (here $d=3$ for the purpose of visualisation). By construction, $J_x(\bm{r})$ thus corresponds to the projection of the current vector field $\bm{J}(\bm{r})$ on the unit vector normal to $\Omega$. The net current is obtained by integrating $J_x(\bm{r})$ over $\bm{r} \in \Omega$.}
    \label{fig:schematic_currents}
\end{figure}

Here we outline the calculation required to compute the asymptotic variance of the integrated current across a mesostate interface, Eq.~\eqref{eq:def_variance_curr}, starting from the spectral density tensor of the probability current vector field, Eq.~\eqref{eq:sp_dens_j}. These arguments closely follow the standard treatment of hyperuniform fluctuations, reviewed in \cite{torquato_hyperuniformity_2016}.
Without loss of generality we assume that the interface, denoted $\Omega$ henceforth, is a $d-1$ dimensional hypersurface of constant $x=0$ coordinate embedded in $d$ dimensional space (see Figure \ref{fig:schematic_currents}). We introduce $\tilde{\bm{r}}$ as the $d-1$ dimensional vector satisfying $\bm{r} = (r_x,\tilde{\bm{r}})$.
The variance of the integrated current across the interface $\Omega$ is thus given by
\begin{align}
\textit{Var}(J;\Omega) &= \int_\Omega d^{d-1}\tilde{r}' \int_\Omega d^{d-1}\tilde{r}'' \ \langle J_x(0,\tilde{\bm{r}}') J_x(0,\tilde{\bm{r}}'') \rangle  \\
&= \int_\Omega d^{d-1}\tilde{r}' \int_\Omega d^{d-1}\tilde{r}'' \ C_{xx}(0,\tilde{\bm{r}}'-\tilde{\bm{r}}'') \\
&= v_\Omega \int d^{d-1}\tilde{r}' \ C_{xx}(0,\tilde{\bm{r}}') \gamma(\tilde{\bm{r}}';\Omega) \label{eq:var_from_cxx}
\end{align}
where
\begin{equation}
\gamma(\tilde{\bm{r}}';\Omega) = \frac{1}{v_\Omega} \int d^{d-1}\tilde{r}'' \ \mathcal{I}_\Omega(\tilde{\bm{r}}'') \mathcal{I}_\Omega(\tilde{\bm{r}}'+\tilde{\bm{r}}'') 
\end{equation}
is the overlap surface fraction, with $\mathcal{I}_\Omega$ the indicator function over the interface, and $v_\Omega$ is the interface surface.
What we require to proceed further is therefore an expression for the component $C_{xx}$ of the current correlation function for displacements on the interface.
To obtain this, we start from the corresponding component of the spectral density, Eq.~\eqref{eq:final_corr_fn_app_der}, and introduce $\tilde{\bm{k}}$ as the $d-1$ dimensional vector satisfying $\bm{k} = (k_x,\tilde{\bm{k}})$ to write 
\begin{equation}
	\langle \mathcal{J}_x(\bm{k}) \mathcal{J}_x(\bm{k}') \rangle  = \langle \mathcal{J}_x(k_x,\tilde{\bm{k}}) \mathcal{J}_x(k_x',\tilde{\bm{k}}') \rangle  =\frac{4 \tilde{\lambda}}{V} \frac{|\tilde{\bm{k}}|^2}{(|\tilde{\bm{k}}|^2 + k_x^2)^{1+\frac{\eta}{2}}} \delta_{\tilde{\bm{k}}+\tilde{\bm{k}}',0} \delta_{k_x+k_x',0}~. \label{eq:cov_appc}
\end{equation}
The correlation function $C_{xx}$ is thus obtained from Eq.~\eqref{eq:cov_appc} by Fourier back-transforming according to the convention Eq.~\eqref{eq:conv_ft}, 
\begin{align}
C_{xx}(0,\tilde{\bm{r}}) &= \langle J_x(0,0) J_x(0,\tilde{\bm{r}})\rangle \\
&= \frac{1}{V^2} \sum_{k_x,\tilde{\bm{k}}} e^{i \tilde{\bm{k}} \cdot \tilde{\bm{r}}} \langle \mathcal{J}_x(k_x,\tilde{\bm{k}}) \mathcal{J}_x(-k_x,-\tilde{\bm{k}}) \rangle \\
&= \int \dbar^{d-1} \tilde{k} \ e^{i \tilde{\bm{k}}\cdot\tilde{\bm{r}}} \int \dbar k_x \ \frac{4\tilde{\lambda}}{V^2} \frac{|\tilde{\bm{k}}|^2}{(|\tilde{\bm{k}}|^2 + k_x^2)^{1+\frac{\eta}{2}}} \\
&= \int \dbar^{d-1} \tilde{k} \ e^{i \tilde{\bm{k}}\cdot\tilde{\bm{r}}} \left[ \frac{2\tilde{\lambda}  \Gamma\left( \frac{1+\eta}{2} \right)}{V^2 \sqrt{\pi} \Gamma\left( 1 + \frac{\eta}{2}\right)} |\tilde{\bm{k}}|^{1-\eta} \right] ~, \label{eq:spectral_density_hyp}
\end{align}
where we have approximated sums over momenta, which involve an infrared cutoff at a typical inverse length-scale $V^{-1/d}$, as integrals in the limit of large $V$ on the basis of the infrared convergence of the integral Eq.~\eqref{eq:spectral_density_hyp} for $\eta < d$. Following the usual convention \cite{zinn2021quantum}, dashed differentials are a short-hand for $\dbar^d k= d^d k/(2\pi)^d$. It is interesting to note that, while the full spectral density Eq.~\eqref{eq:sp_dens_j} is not hyperuniform for $\eta=0$, the relevant component of the corresponding correlation function is hyperuniform for $\eta<1$ with hyperuniformity exponent $1-\eta$ when we concentrate on the lower-dimensional object that is the mesostate interface. 
For the correlation function $C_{xx}(0,\tilde{\bm{r}})$ origination from Eq.~\eqref{eq:spectral_density_hyp} to be well-defined in the range $\eta<d$, the Fourier back-transform needs to be regularised in the ultraviolet. Here, we do so by introducing a soft cutoff, suggestive of the introduction of a microscopic lattice spacing of typical length scale $\beta = \ell$. Denoting $C^R_{xx}(0,\tilde{\bm{r}};\beta)$ the regularised correlation function, we write
\begin{align}
C^R_{xx}(0,\tilde{\bm{r}}; \beta) &= \int \dbar^{d-1} \tilde{k} \ e^{i \tilde{\bm{k}}.\tilde{\bm{r}}} \left[ \frac{2\tilde{\lambda} \Gamma\left( \frac{1+\eta}{2} \right)}{V^2 \sqrt{\pi} \Gamma\left( 1 + \frac{\eta}{2}\right)} |\tilde{\bm{k}}|^{1-\eta} \right] e^{-\beta |\tilde{\bm{k}}|} \label{eq:cyl_int}\\
&= \mathcal{N}(\eta,d;\tilde{\lambda},V)  \int_0^{\pi} d\theta \ (\sin(\theta))^{d-3}\int_0^\infty d\tilde{k} \ e^{i \tilde{k} \cos(\theta) |\bm{\tilde{r}}| - \beta \tilde{k}} \ \tilde{k}^{d-1-\eta} \label{eq:cyl_int_redux} \\
&\propto \sum_{n=0}^\infty \frac{(-1)^n}{(2n)!} |\bm{\tilde{r}}|^{2n} \frac{\Gamma\left(\frac{d}{2}-1\right)\Gamma\left(n+\frac{1}{2}\right)}{\Gamma\left(n+\frac{d}{2}-1\right)} \int_0^\infty d\tilde{k} \ e^{-\beta \tilde{k}} \tilde{k}^{d-1-\eta+2n} \label{eq:theta_int_interm}\\
&\propto \sqrt{\pi} \frac{\Gamma\left( d-\eta \right) \Gamma\left( \frac{d-2}{2} \right)}{\Gamma\left( \frac{d-1}{2} \right)} \beta^{-d+\eta} \prescript{}{2}{F}_1\left( \frac{d-\eta}{2}, \frac{d-\eta+1}{2},\frac{d-1}{2}; - \frac{|\bm{\tilde{r}}|^2}{\beta^2} \right) 
\label{eq:regul_corr} ~.
\end{align}
for $\eta < d$ and $d>2$, where $\prescript{}{2}{F}_1(a,b,c;z)$ denotes the hypergeometric function \cite{bateman}. We note for later use that $\prescript{}{2}{F}_1(a,b,c;z)$ has the asymptotic expansion
\begin{equation}
    \prescript{}{2}{F}_1(a,b,c;z) = \lambda_1 z^{-a} + \lambda_2 z^{-b} + \mathcal{O}(z^{-a-1},z^{-b-1}) \label{eq:asympt_sypergeom}
\end{equation}
with 
\begin{equation}
    \lambda_1 = \frac{\sqrt{\pi} \Gamma\left( \frac{d-1}{2}\right)}{\Gamma \left(\frac{\eta -1}{2}\right) \Gamma \left(\frac{d+1-\eta}{2}\right)}
\end{equation}
at large $|z|$ unless $a-b$ is an integer \cite{bateman}.
To go from Eq.~\eqref{eq:cyl_int} to \eqref{eq:cyl_int_redux} we first perform a change of variable to spherical coordinates using
\begin{equation}
    \int d^n \tilde{k} = \int_0^\infty d\tilde{k} \int_0^\pi d\theta d\phi_1 ... d\phi_{n-3} \int_0^{2\pi} d\phi_{n-2} \ \tilde{k}^{n-1} \sin^{n-2}(\theta) \sin^{n-3}(\phi_1) ... \sin(\phi_{n-3})
\end{equation}
for $n=d-1$ and $n>1$, with $\tilde{k} = |\tilde{\bm{k}}|$ the wavenumber and $\theta = \arccos[(\tilde{\bm{k}}/|\tilde{\bm{k}}|) \cdot (\tilde{\bm{r}}/ |\tilde{\bm{r}}|)]$ the angle between the wavevector $\tilde{\bm{k}}$ and the displacement $\tilde{\bm{r}}$. Since the integrand of Eq.~\eqref{eq:cyl_int} is a function only of $\tilde{k}$ and $\theta$, the remaining angular coordinates can be integrated out. To go from Eq.~\eqref{eq:cyl_int_redux} to \eqref{eq:theta_int_interm} we perform a second change of variable $u(\theta) = \cos(\theta)$ and expand the ensuing trigonometric functions as power series in $\tilde{k} u |\bm{\tilde{r}}|$ before carrying out the integral over $u$.
In the following, we shall ignore the numerical prefactor
\begin{equation}
\label{eq:prefactor_collect_var}
    \mathcal{N}(\eta,d;\tilde{\lambda},V) = \frac{1}{(2\pi)^{d-1}} \frac{2\tilde{\lambda} \Gamma\left( \frac{1+\eta}{2} \right)}{V^2 \sqrt{\pi}  \Gamma\left( 1 + \frac{\eta}{2}\right)} \frac{2 \pi^\frac{d-2}{2}}{\Gamma\left(\frac{d-2}{2}\right)}
\end{equation}
in Eq.~\eqref{eq:cyl_int_redux}, since is independent of both $\theta$ and $k$ and thus irrelevant for the window size scaling. The case $d=2$ needs to be treated separately and produces for $\eta<d$:
\begin{equation}
    C^R_{xx}(0,\tilde{\bm{r}}; \beta) \propto \Gamma(2-\eta) \beta^{-2+\eta}  \left( 1+\frac{|\tilde{\bm{r}}|^2}{\beta^2} \right)^{\frac{\eta-2}{2}} \cos\left[(2-\eta) \arctan\left(\frac{|\tilde{\bm{r}}|}{\beta}\right)\right]~,
\end{equation}
which is characterised by the same leading order asymptotic scaling with $|\tilde{\bm{r}}|$ at $|\tilde{\bm{r}}|/\beta\gg 1$ as Eq.~\eqref{eq:regul_corr} for $d \to 2$.

Taking the interface as a hypersphere of radius $\mathcal{L}=L\ell$, whence $v_\Omega \sim \mathcal{L}^{d-1}$, it was shown \cite{torquato_hyperuniformity_2016} that, for $|\tilde{\bm{r}}|<2\mathcal{L}$,
\begin{equation}
\gamma(\tilde{\bm{r}};\Omega) =1 - c(d) \left(\frac{|\tilde{\bm{r}}|}{\mathcal{L}}\right) + c(d) \sum_{n=2}^\infty (-1)^n \frac{(d-1)(d-3)...(d-2n+3)}{(2n-1)! [2\cdot 4 \cdot6 ... (2n-2)]} \left(\frac{|\tilde{\bm{r}}|}{\mathcal{L}}\right)^{2n-1}
\end{equation}
with $c(d)= 2 \Gamma(1+d/2)/ (\sqrt{\pi} \Gamma((d+1)/2))$. Ref.~\cite{kim_effect_2017} explores the effect of changing the interface shape. Overall 
\begin{equation}
\textit{Var}(J;\Omega) = v_\Omega  \int_\Omega d^{d-1}\tilde{r} \ C_{xx}(0,\tilde{\bm{r}}) \left[ 1 - c(d) \left(\frac{|\tilde{\bm{r}}|}{\mathcal{L}}\right) + c(d) \sum_{n=2}^\infty (-1)^n \frac{(d-1)(d-3)...(d-2n+3)}{(2n-1)! [2 \cdot 4 \cdot 6 ... (2n-2)]} \left(\frac{|\tilde{\bm{r}}|}{\mathcal{L}}\right)^{2n-1} \right]~. \label{eq:expanded_var}
\end{equation}
The term in Eq.~\eqref{eq:expanded_var} originating from the zeroth order in the series expansion of $\gamma$ approaches, in the asymptotic limit of $\mathcal{L} \to \infty$, the value of the spectral density, Eq.~\eqref{eq:spectral_density_hyp}, at $\tilde{\bm{k}}=0$,
\begin{equation}
    \lim_{\mathcal{L} \to \infty} \int_\Omega d^{d-1}\tilde{r} \ C_{xx}(0,\tilde{\bm{r}}) = \lim_{\tilde{\bm{k}}\to0} \left[  \frac{2\tilde{\lambda}  \Gamma\left( \frac{1+\eta}{2} \right)}{V^2 \sqrt{\pi} \Gamma\left( 1 + \frac{\eta}{2}\right)} |\tilde{\bm{k}}|^{1-\eta} \right] \label{eq:cancell_hypun}
\end{equation}
and thus vanishes when the latter is hyperuniform, namely for $\eta < 1$.

Based on the expansion Eq.~\eqref{eq:asympt_sypergeom}, the behaviour of the correlation function, Eq.~\eqref{eq:regul_corr}, for $|\tilde{\bm{r}}|/\beta \gg 1$ is $C^R_{xx}(0,\tilde{\bm{r}}; \beta) \sim |\tilde{\bm{r}}|^{\eta-d}$ therefore each term in the integrand in the right-hand side of Eq.~\eqref{eq:expanded_var} behaves like $|\tilde{\bm{r}}|^{\eta-2 + m}$ with $m \in \{0,1,3,...\}$. 
The scaling of the variance with window size $\mathcal{L}$ is thus controlled by the infrared divergence as $\mathcal{L} \to \infty$. For $\eta < 1- m $, the integrals are infrared convergent, which allows us to let $\Omega \to \mathbb{R}^{d-1}$ without modifying the leading order asymptotic scaling with $\mathcal{L}$.
Proceeding case by case:
\begin{itemize}
\item
$\bm{0 < \eta < d}$:  For $\eta > 1$, the relevant integrals are infrared divergent. For $0<\eta<1$ the term originating from the zeroth order in the expansion of $\gamma$ is removed by the cancellation Eq.~\eqref{eq:cancell_hypun}.
For $\eta = 1$, based on Eq.~\eqref{eq:spectral_density_hyp}, the correlation function $C_{xx}$ takes the form of a Dirac delta in real space so that the term originating from the zeroth order in the expansion of $\gamma$ is finite but $\mathcal{L}$ independent and does not contribute to the scaling.
Overall, the scaling is dominated by the infrared divergent behaviour at large $|\tilde{\bm{r}}|$. Substituting Eq.~\eqref{eq:regul_corr} together with the asymptotic expansion \eqref{eq:asympt_sypergeom} into Eq.~\eqref{eq:expanded_var} and performing a change of variable $\tilde{\bm{r}} \to \tilde{\bm{r}}/\mathcal{L}$ to isolate the $\mathcal{L}$ dependence, we arrive at $\textit{Var}(J;\Omega) \sim \mathcal{L}^{d-2+\eta}$ for $0 < \eta < d$. 
\item 
$\bm{\eta=0}$: The case of $\eta = 0$ needs to be treated separately because the integral of the first order term in the expansion of $\gamma(\tilde{\bm{r}};\Omega)$ is logarithmically divergent.
Substituting Eq.~\eqref{eq:regul_corr} for the regularised correlation function into Eq.~\eqref{eq:expanded_var} and setting $\eta=0$ we are left with integrals of the form
\begin{align}
    \mathcal{I}_m(\mathcal{L}) 
    &= v_\Omega \int_\Omega d^{d-1}\tilde{r} \
    \prescript{}{2}{F}_1\left( \frac{d}{2}, \frac{d+1}{2},\frac{d-1}{2}; - \frac{|\bm{\tilde{r}}|^2}{\beta^2} \right) 
    \left( \frac{|\tilde{\bm{r}}|}{\mathcal{L}} \right)^m ~.
\end{align}
for $m \geq 1$. For $m>1$ and based on the expansion Eq.~\eqref{eq:asympt_sypergeom}, the $\tilde{\bm{r}}$ integral is dominated by the behaviour of the integrand in the large $|\tilde{\bm{r}}|$ regime and we recover the $\mathcal{L}^{d-2}$ scaling obtained upon setting $\eta=0$ in the scaling law for the case $0<\eta<d$.
For $m=1$ we exploit the spherical symmetry of the integral to write
\begin{equation}
    \int_\Omega d^{d-1}\tilde{r} \ f(|\tilde{\bm{r}}|) = \int_0^\mathcal{L} d|\tilde{\bm{r}}| \ \frac{2\pi^{\frac{d-1}{2}}}{\Gamma(\frac{d-1}{2})} |\tilde{\bm{r}}|^{d-2} f(|\tilde{\bm{r}}|)
\end{equation}
and invoke the asymptotic expansion Eq.~\eqref{eq:asympt_sypergeom} to obtain the leading order behaviour
\begin{equation}
    \mathcal{I}_{m=1}(\mathcal{L}) = \frac{v_\Omega}{\mathcal{L}} \int_0^\mathcal{L} \frac{d|\tilde{\bm{r}}|}{\beta} \ \lambda_1 \frac{2\pi^{\frac{d-1}{2}} \Gamma(d)}{ \Gamma(\frac{d-1}{2})}  \left( \frac{|\tilde{\bm{r}}|}{\beta} \right)^{-1} \sim \mathcal{L}^{d-2} \log(\mathcal{L}/\beta) \label{eq:log_scal_L}~,
\end{equation}
which dominates the asymptotic scaling with $\mathcal{L}$. The variance of the integrated current thus scales like $\textit{Var}(J;\Omega)\sim \mathcal{L}^{d-2}\log(\mathcal{L}/\beta)$ for $\eta = 0$.
\item
$\bm{\eta<0}$: Finally, we consider the case of $\eta < 0$. Substituting once again Eq.~\eqref{eq:regul_corr} into Eq.~\eqref{eq:expanded_var}, we obtain
\begin{equation}
    \mathcal{I}_m(\mathcal{L}) = 
    v_\Omega \int_\Omega d^{d-1}\tilde{r} \
    \prescript{}{2}{F}_1\left( \frac{d-\eta}{2}, \frac{d-\eta+1}{2},\frac{d-1}{2}; - \frac{|\bm{\tilde{r}}|^2}{\beta^2} \right) 
    \left( \frac{|\tilde{\bm{r}}|}{\mathcal{L}} \right)^m ~. 
\end{equation}
for $m \in \{1,3,5,...\}$. When $m < 1-\eta$ this integral is infrared convergent and we can take $\Omega \to \mathbb{R}^{d-1}$ without affecting the leading order scaling with $\mathcal{L}$. Upon taking this limit, the $\mathcal{L}$ dependence is limited to the prefactors and we straightforwardly obtain $\mathcal{I}_m(\mathcal{L}) \sim \mathcal{L}^{d-1-m}$. When $m=1-\eta $ the integral is logarithmically divergent and we obtain $\mathcal{I}_m(\mathcal{L}) \sim \log(\mathcal{L}/\beta) \mathcal{L}^{d-2+\eta}$ along the lines of Eq.~\eqref{eq:log_scal_L}. Finally, when $m>1-\eta $, the integral is infrared divergent, whereby we invoke the asymptotic expansion, Eq.~\eqref{eq:asympt_sypergeom}, to obtain $\mathcal{I}_m(\mathcal{L}) \sim \mathcal{L}^{d-2+\eta}$.
Overall, the  scaling is dominated by the $m=1$ term and follows $\textit{Var}(J;\Omega)\sim \mathcal{L}^{d-2}$ for $\eta < 0$.
\end{itemize}
Combining the results that we just derived into a single expression and making the dependence on the model parameters $\tilde{\lambda}$ and $V$ appearing in the prefactor Eq.~\eqref{eq:prefactor_collect_var} explicit, we thus have
\begin{equation}
    \textit{Var}(J;\Omega) \sim \left\{\begin{array}{lr}
        \tilde{\lambda} V^{-2} \mathcal{L}^{d-2} , & \text{for } \eta < 0\\
        \tilde{\lambda} V^{-2} \mathcal{L}^{d-2}\ln(\mathcal{L}/\beta) , & \text{for } \eta = 0\\
        \tilde{\lambda} V^{-2} \beta^{-\eta} \mathcal{L}^{d-2+\eta} , & \text{for } 0<\eta<d
        \end{array} \right.~.
\end{equation}
Substituting into Eq.~\eqref{eq:scal_si_mes} with $\beta=\ell$,
\begin{equation}
    \left\langle \dot{s}_i^{\rm (meso)} \right\rangle(\mathcal{L}) \sim \left\{\begin{array}{lr}
        \tilde{\lambda} \ell^3  V^{-1} h^{-1} \mathcal{L}^{-1}, & \text{for } \eta < 0\\
        \tilde{\lambda} \ell^3  V^{-1} h^{-1} \mathcal{L}^{-1}\ln(\mathcal{L}/\ell), & \text{for } \eta = 0\\
        \tilde{\lambda} \ell^{3-\eta} V^{-1} h^{-1} \mathcal{L}^{-1+\eta}, & \text{for } 0<\eta<d
        \end{array} \right.~,
\end{equation}
which constitutes our main result.

\bibliography{en_scaling_refs}

\makeatletter\@input{xx.tex}\makeatother